\documentclass[lettersize,journal]{IEEEtran}
\usepackage{amsmath,amsfonts,amssymb}
\usepackage{graphicx}
\usepackage{subfig}
\usepackage{algorithmicx}
\usepackage{algorithm}
\usepackage{algpseudocode}
\usepackage{textcomp}
\usepackage{multirow,multicol}
\usepackage[colorlinks]{hyperref}
\hypersetup{citecolor=blue}
\usepackage[table]{xcolor}
\usepackage{threeparttable}
\usepackage{balance}
\usepackage{booktabs}
\usepackage[mathscr]{eucal}
\newtheorem{remark}{Remark}
\newtheorem{theorem}{Theorem}
\newtheorem{corollary}{Corollary}
\newtheorem{lemma}{Lemma}

\usepackage[font=small,skip=2pt]{caption}

\usepackage{enumitem}
\setlist{nosep}

\begin{document}
\title{Finite-blocklength Fluid Antenna Systems}
\author{Zhentian Zhang,~\IEEEmembership{Member,~IEEE}, 
Kai-Kit Wong,~\IEEEmembership{Fellow,~IEEE}, 
\\David Morales-Jimenez,~\IEEEmembership{Senior Member,~IEEE},
Hao Jiang,~\IEEEmembership{Senior Member,~IEEE}, 
Hao Xu,~\IEEEmembership{Member,~IEEE}, 
Christos Masouros,~\IEEEmembership{Fellow,~IEEE,}
and Zaichen Zhang,~\IEEEmembership{Senior Member,~IEEE}
\vspace{-10mm}
\thanks{The work of Kai-Kit Wong is partly supported by the Engineering and Physical Sciences Research Council (EPSRC) under Grant EP/W026813/1. The work of Hao Jiang is partly supported in part by the National Natural Science Foundation of China Projects under Grant 62471238. The work of Zaichen Zhang is partly supported by the Fundamental Research Funds for the Central Universities 2242022k60001, and the research fund of National Mobile Communications Research Lab. 2026A03. The work of Hao Xu is partly supported by the National Natural Science Foundation of China under Grant U25A20398. The work of D. Morales-Jimenez is partly supported in part by the AEI of Spain and the European Social Fund under Grant RYC2020-030536-I, and in part by MICIU/AEI/10.13039/501100011033 and FEDER/UE under Grant PID2023-149975OB-I00 (COSTUME).}
\thanks{Zhentian Zhang is with the National Mobile Communications Research Laboratory, Southeast University, Nanjing, 210096, China and he is also with the Hong Kong Polytechnic University, Hong Kong SAR, China. (e-mail: zhentianzhangzzt@gmail.com).}
\thanks{Zaichen Zhang is with the National Mobile Communications Research Laboratory, Frontiers Science Center for Mobile Information Communication and Security, Southeast University, Nanjing, 210096, China and he is also with the Purple Mountain Laboratories, Nanjing 211111, China (e-mail: zczhang@seu.edu.cn).}
\thanks{David Morales-Jimenez is with the Department of Signal Theory, Networking and Communications, University of Granada, Granada 18071, Spain (e-mail: dmorales@ugr.es).}
\thanks{Kai-Kit Wong and Christos Masouros are with the Department of Electronic and Electrical Engineering, University College London, Torrington Place, WC1E 7JE, United Kingdom. Kai-Kit Wong is also affiliated with the Department of Electronic Engineering, Kyung Hee University, Yongin-si, Gyeonggi-do 17104, Korea.  (e-mails: \{kai-kit.wong, c.masouros\}@ucl.ac.uk).}
\thanks{Hao Jiang is with the School of Cyber Science and Engineering, Southeast University, Nanjing 210096, P. R. China (e-mail: jiang.hao@seu.edu.cn).}
\thanks{Hao Xu is with the National Mobile Communications Research Laboratory, Southeast University, Nanjing 210096, China. He is also with the School of Intelligent Science and Engineering, Southeast University Wuxi Campus, Wuxi 214061, China (e-mail: hao.xu@seu.edu.cn).}
\thanks{{\em Corresponding authors: Kai-Kit Wong, Hao Jiang.}}
}

\maketitle

\begin{abstract}
This paper investigates fluid antenna systems (FASs) subject to finite-blocklength (FBL) constraints, motivated by the strict reliability-latency and ultra-massive connectivity requirements of future wireless networks. While FAS performance has been widely studied in the asymptotic regime, its behavior under FBL remains largely unexplored. Our objective is to develop a unified set of analytical tools for evaluating FASs under FBL that remains applicable across different spatial-correlation models. First, to establish accurate benchmarks for non-orthogonal finite-length user signature design, we characterize both the average and the worst-case correlation coefficients via extreme value theory (EVT) and derive closed-form predictions of the achievable correlation levels. Second, taking block error rate (BLER) as the fundamental FBL metric, we study joint detection and decoding in FAS-assisted links and derive a closed-form BLER expression that is universally applicable across channel models. Additionally, we revisit outage probability (OP) in the FBL regime and obtain tractable OP characterizations for both FASs and conventional multiple fixed-position antenna (FPA) systems. In order to reduce the computational burden for multi-fold integrals in correlated fading models, we further propose a Taylor-expansion-assisted mean value theorem for integrals (MVTI), thus enabling efficient performance evaluation with marginal accuracy loss. Numerical results validate the analysis and reveal that even single-antenna FASs can have superior spatial diversity relative to conventional multi-FPA systems. Moreover, under both FBL and interference-limited environments, FASs provide improved energy, spectral, and hardware efficiencies, hence highlighting FAS as a promising enabler for next-generation wireless networks.
\end{abstract}

\begin{IEEEkeywords}
Fluid antenna system, finite blocklength, codeword cross-correlation, block error rate, outage probability.
\end{IEEEkeywords}

\section{Introduction}
\subsection{Background and Related Work}
\IEEEPARstart{S}{ixth}-generation (6G) and beyond networks \cite{6G1} envision essential use scenarios, including immersive communication, hyper-reliable low-latency communication (HRLLC), integrated sensing and communication (ISAC), ubiquitous communication, massive connectivity, and artificial intelligence. To meet these stringent requirements, one of the most critical design aspects for current and future systems lies in operation within the finite blocklength (FBL) regime \cite{intro_FBL1,ML_URA}. Accordingly, multiple-antenna systems \cite{intro_FBL2} become essential due to the rich spatial diversity \cite{intro_FBL3,FBL_MIMO}. Though the increase in array size is effective for capacity enhancement, it suffers from poor scalability due to signal processing overhead and hardware cost. Thus, there is strong desire to raise the system performance without increasing the antenna count.

This is addressed by the emerging concept, known as the fluid antenna system (FAS) \cite{fas-twc-21} which is a hardware-agnostic system concept that regards the antenna as a reconfigurable physical-layer resource for system optimization. An important feature of FAS is the utilization of position reconfigurability for diversity gains even within a compact aperture \cite{fas3}. Since its introduction, it has been investigated in a wide range of scenarios for great effects \cite{fas_tutorial,FAS_survey_Hong,Lu-2025,FAS_enabler,FAS_wu_tuo1,AD}. Selected examples include channel modeling for diversity analysis \cite{kit_electronic,block1,new_fas}, synergy between FAS and intelligent surfaces \cite{block2}, FAS for unmanned aerial vehicules (UAVs) \cite{fas2}, FAS-assisted ISAC \cite{chernoff3}, and fluid antenna multiple access (FAMA) \cite{FAMA1,fama-overview2026,chernoff2}, to name a few. The FAS concept has also inspired new reconfigurable antenna technologies such as liquid antennas \cite{shen2024design,Shamim-2025}, metamaterial-based antennas \cite{Zhang-jsac2026,Liu-2025arxiv} and reconfigurable pixel antennas \cite{zhang2024pixel,tong-2025pixel,Wong-wc2026}, etc. Moreover, dual-sided FASs with joint correlated channel modeling is first investigated in \cite{dual_side}.

Beyond the widely studied fading-diversity gains of single-port fluid antennas, fluid antenna arrays (FAAs) offer a complementary and potentially more fundamental advantage: \emph{geometry diversity} \cite{FAA1,FAA2}. Since this capability arises from deliberate reconfiguration of the array geometry, it can be exploited with little additional overhead. By reshaping the effective aperture, spatial sampling structure, and spatial-frequency response, an FAA enables radiation-pattern control beyond that achievable with fixed arrays. As demonstrated for both linear and planar FAAs in \cite{FAA1,FAA2}, such geometric flexibility supports enhanced beamforming gain, directivity shaping, and, most importantly for interference-limited systems, sidelobe suppression \cite{FAA3,FAA4}. Worth noting that, mutual coupling effect for FAAs is investigated in \cite{FAA5}.

\subsection{Challenges and Motivation}
Despite the growing body of FAS literature, a rigorous and systematic performance characterization under FBL remains largely unexplored. Though recent efforts have begun to study FAS in the FBL regime \cite{block_FBL_FAS,FBL_FAS2}, the studies are limited to the single-user case. To extend the work to multiuser settings, several fundamental challenges arise:
\begin{itemize}
\item {\bf Lack of theoretical benchmark for non-orthogonal signature design}---Under FBL constraints, orthogonality-based signatures do not scale well. Thus, non-orthogonal signature design becomes important for improving the network spectral efficiency \cite{codeword_design1,codeword_design2}. Although all non-orthogonal signature designs should asymptotically approach orthogonality, establishing solid theoretical benchmarks to evaluate the correlation of finite-length signatures/codewords remains an open problem.
\item {\bf Lack of universal performance metric for FAS under FBL}---Most existing FAS-based researches focus on tail-behavior analysis in the asymptotic regime based on specific distributions of the post-selection channel gain. Such analyses rely heavily on channel modeling, meaning that no further insight can be obtained without accurate analytical characterization of channel correlation. In particular, the analysis becomes even more complicated when FBL constraints are taken into account.
\end{itemize}

These challenges motivate this work to provide fundamental analytical solutions to the above problems, with a particular focus on completing the performance analysis of FAS in the non-asymptotic regime (i.e., FBL). Specifically, this work aims to accurately characterize the correlation of non-orthogonal signatures and to develop universal performance metrics with the potential to accommodate various channel models.

\subsection{Contributions}
This work presents an analytical framework for FAS under FBL and quasi-static fading, with particular emphasis on \emph{codeword cross-correlation characterization} and \emph{universal performance metrics}, including the block error rate (BLER) and the redefined outage probability (OP). The proposed closed-form expressions for BLER and theoretical non-orthogonal correlation are universally applicable. The main contributions of this paper are summarized as follows:
\begin{itemize}
\item \textbf{Analytical prediction on signature/codewords correlation}---To address the lack of a universally applicable characterization of signature/codeword correlation, we analyze the average and maximum correlation of finite-length codewords. Closed-form expressions are derived based on extreme value theory (EVT), enabling accurate prediction of correlation-induced interference.
\item \textbf{Closed-form BLER expression}---Unlike existing BLER analyses that rely on Gaussian approximations \cite{block_FBL_FAS,FBL_FAS2} and require exact signal-to-interference plus noise ratio (SINR) evaluation, we formulate a block-wise maximum likelihood (ML) detection model to realize \emph{joint detection and decoding}, and derive a closed-form upper bound on BLER conditioned on the post-selection effective gain. Our expression is applicable to different correlation characterizations in FAS, based on either the instantaneous channel response or its distribution.
\item \textbf{Redefining OP and analysis for FBL-FAS}---In contrast to the prevailing asymptotic analyses based on the infinite blocklength assumption \cite{fas-twc-21,fas3,fas_tutorial}, we redefine the OP for FAS under the FBL constraints as the event that {\em the instantaneous mutual information falls below the target coding rate} \cite{FBL_MIMO}. Based on this definition, tractable OP expressions are derived. In addition, a closed-form OP expression is also obtained for conventional multi-FPA systems employing maximal ratio combining (MRC).
\item \textbf{Multi-dimensional integral approximation}---To alleviate the high computational complexity caused by the product-integral probability density functions arising even in simple reference channel correlation models, we propose a simplification method termed Taylor-expansion-assisted Mean Value Theorem for Integrals (MVTI). The proposed approximation substantially simplifies the analysis while incurring negligible accuracy loss.
\end{itemize}

\textit{Notations:} Scalars are denoted by lowercase and uppercase letters (e.g., $a$, $A$), and vectors and matrices by bold lowercase and uppercase letters (e.g., $\boldsymbol{a}$, $\boldsymbol{A}$), respectively. $\mathbb{R}$ and $\mathbb{C}$ denote the real and complex fields. Calligraphic letters (e.g., $\mathcal{A}$) denote sets, and $\mathcal{A}\setminus\mathcal{A}'$ denotes set difference. $\mathrm{E}[\cdot]$, $\|\cdot\|_2^2$, $(\cdot)^*$, $(\cdot)^{\mathrm{T}}$, and $(\cdot)^{\mathrm{H}}$ denote expectation, squared $\ell_2$-norm, conjugate, transpose, and Hermitian transpose, respectively. $\mathcal{G}$, $\mathcal{CN}$, $\mathscr{G}_z$, and $\mathcal{E}$ denote the Gamma, circularly symmetric complex Gaussian, Gumbel, and exponential distributions, respectively. Unless otherwise stated, $\log(\cdot)$ is the natural logarithm.

\section{System Model}\label{sec:system model}
Consider a single-FPA base station serving $U$ users, each with a FAS in the uplink. Each users' antenna can be dynamically switched among $N$ predefined locations, which are evenly distributed along a linear aperture of length $W\lambda$, in which $\lambda$ denotes the carrier wavelength and $W$ denotes the antenna aperture length constant normalized by wavelength. Each antenna at a given location is referred to as a {\em port} and modeled as an ideal point antenna. Quasi-static fading is assumed throughout.

\subsection{Different Channel Correlation Models}\label{sec.models}
{\em Simple Reference Channel Model}---In the earliest results, the first port is usually taken as the reference location, with the displacement of the $k$-th port measured relative to it as
\begin{equation}\label{eq:1}
\Delta d_{k,1} = \left(\frac{k-1}{N-1}\right)W\lambda,~k=1,2,\ldots,N.
\end{equation}
Let $g_{u,k}\sim{\cal CN}(0,\sigma^2)$ denote the channel coefficient of the $u$-th user at the $k$-th port. Under this model, the channel response amplitude $\rvert g_{u,k} \lvert$ is Rayleigh distributed with probability density function (PDF)
\begin{equation}\label{eq:2}
p_{\rvert g_{u,k} \lvert} \left(r\right) = \frac{2r}{\sigma^2}e^{-\frac{r^2}{\sigma^2}},~\text{for}~r\ge 0~\text{with}~\mathrm{E}\left[\rvert g_{u,k} \lvert^2\right]=\sigma^2.
\end{equation}
The channels at all $N$ ports are written as \cite{fas-twc-21}
\begin{equation}\label{eq:3}
	\left\{\begin{matrix}
		\begin{aligned}
		g_{u,1}&=\sigma x_{u,0}+j\sigma y_{u,0}\\
		g_{u,k}&=\sigma\left(\sqrt{1-\mu_{k}^{2}}x_{u,k}+\mu_{k}x_{u,0}\right)+\\
		&j\sigma\left(\sqrt{1-\mu_{k}^{2}}y_{u,k}+\mu_{k}y_{u,0}\right)~\mathrm{for}~k=2,\ldots,N,
		\end{aligned}
	\end{matrix}\right.
\end{equation}
where $x_{u,0},x_{u,2},\ldots,x_{u,N},y_{u,0},y_{u,2},\ldots,y_{u,N}$ are all independent real Gaussian random variables with zero mean and variance $\frac{1}{2}$, and $\left\{\mu_k\right\}$ are parameters quantifying the ports' correlation with respect to the first port:
\begin{equation}\label{eq:4}
	\mu_k=J_0\left(\frac{2\pi\left(k-1\right)}{N-1}W\right),~\text{for}~k=1,2,\ldots,N,
\end{equation} 
 where $J_0\left(\cdot\right)$ is the zero-order Bessel function of the first kind and $\mu_1=J_0\left(0\right)=1$. The channels $\left\{g_{u,k}\right\}$ are correlated since the ports can be arbitrarily close to each other.

{\em Modified Reference Correlation Model}---The correlation model in \eqref{eq:3}, as an early attempt to characterize FAS correlation, has been widely adopted but lacks accuracy because it only considers correlation relative to a reference port and fails to capture the relationship between arbitrary adjacent ports. To address this limitation, a modified reference model was introduced in \cite[(3)]{fas_tutorial}, \cite{kit_electronic}, where the correlation constants in \eqref{eq:3} are replaced with a modified correlation parameter:
 \begin{equation}\label{eq:modified}
\mu_k = \mu = \sqrt{2}\sqrt{\,_1F_2\!\left(\tfrac{1}{2};1,\tfrac{3}{2};-\pi^2W^2\right) - \frac{J_1(2\pi W)}{2\pi W}},
 \end{equation}
 where $_1F_2(\cdot;\cdot;\cdot)$ is the generalized hypergeometric function and $J_1(\cdot)$ denotes the first-order Bessel function of the first kind. This model connects all ports within the fluid antenna without relying on a reference port, effectively approximating the average squared spatial correlation of a practical fluid antenna array while maintaining analytical tractability.

{\em Fully Correlated Channel Model}---A fully correlated channel model provides the most accurate description of inter-port correlation based on the channel covariance \cite{fas_tutorial}. Although such modeling guarantees accuracy, it lacks analytical tractability due to complex statistical dependencies. Following the construction method in \cite{block_FBL_FAS}, let $\boldsymbol{g}_u \in \mathbb{C}^{N}$ represent the channel coefficient vector of the $u$-th user, and let $\boldsymbol{\Sigma} \in \mathbb{C}^{N \times N}$ denote the corresponding spatial correlation matrix: 
 \begin{equation}\label{eq:full}
 	\boldsymbol{\Sigma} =
 	\begin{pmatrix}
 		a(0) & a(2) & \dots & a(N-1)\\
 		a(-1) & a(1) & \dots & a(N-2)\\
 		\vdots & \ddots & \vdots & \vdots\\
 		a(-N+1) & \dots & & a(0)
 	\end{pmatrix},
 \end{equation}
 which has a Toeplitz structure and we have
 \begin{equation}
a(n) = \operatorname{sinc}\!\left(\frac{2\pi n W}{N-1}\right).
 \end{equation}
 Given $\boldsymbol{\Sigma}$, the channel vector $\boldsymbol{g}_u$ can be efficiently generated using the eigenvalue-based construction~\cite[(5)]{fas_tutorial}:
 \begin{equation}\label{eq:channel_eigenvalue}
 	\boldsymbol{g}_u = \boldsymbol{Q}\boldsymbol{\Lambda}^{\frac{1}{2}}\boldsymbol{g}_{0},
 \end{equation}
 where $\boldsymbol{Q}$ is the eigenvector matrix from the decomposition $\boldsymbol{\Sigma} = \boldsymbol{Q}\boldsymbol{\Lambda}\boldsymbol{Q}^{\mathrm{H}}$, and $\boldsymbol{g}_{0} \in \mathbb{C}^{N} \sim \mathcal{CN}\!\left(\boldsymbol{0}, \sigma^2\boldsymbol{I}\right)$.

\subsection{FAS-FBL Signal Model}
 Assuming $U$ active users transmitting $M$-length signals synchronously, the received signal is written as
 \begin{equation}\label{eq:5}
\boldsymbol{y}= g_{i,k}\boldsymbol{x}_i+\sum_{u\neq i}^{U}g_{u,k}\boldsymbol{x}_u+\boldsymbol{\eta},
 \end{equation}
 where $\boldsymbol{\eta}\sim{\cal CN}({\bf 0},\frac{\sigma^2_{\eta}}{M}\boldsymbol{I}_M)$ is the additive noise and the signals carrying information $\boldsymbol{x}_u\in \mathbb{C}^{M}$ are generalized as Gaussian codewords with zero mean and variance $\frac{1}{M}$. Our aim is to study how the finite blocklength $M$ and the user density $U$ affect the BLER performance and OP considering codewords correlation. Moreover, the signal-to-noise ratio (SNR) is 
\begin{equation}
\mathrm{SNR} = \frac{\mathrm{E}\left[\|g_{i,k}\boldsymbol{x}_i\|_2^2\right]}{\mathrm{E}\left[\|\boldsymbol{\eta}\|_2^2\right]}=\frac{\sigma^2}{\sigma^2_{\eta}}.
\end{equation}
For recent development in efficient channel reconstruction for FASs, please refer to \cite{CSI1,CSI2,CSI3} where reconstruction limits and practical algorithms are discussed under different spatial covariance priors.
\section{Main Results}\label{Main Results}
In this section, the performance limits of FAS in the FBL regime are investigated in terms of BLER and OP. First, the average and maximum levels of cross-correlation among information codewords are given from Theorem~\ref{Theorem 1} to Theorem~\ref{Theorem 4}. Next, the BLER for joint detection and decoding is derived and explained from Theorem~\ref{Theorem 5} to Theorem~\ref{Theorem 8}. Finally, the OP under the FBL regime is defined for FAS and elaborated in Theorem~\ref{Theorem 9} and Theorem~\ref{Theorem 10}, where the cross-correlation prediction plays a crucial role in the SINR calculation. 

Notably, Theorem~\ref{Theorem 5} derives a closed-form BLER expression conditioned on the post-selection channel response, with all other parameters determined solely by the system configuration. Therefore, the average BLER can be evaluated either empirically or theoretically under any channel correlation models, as further supported by Theorem~\ref{Theorem 8}. Similarly, Theorem~\ref{Theorem 9} establishes an outage threshold that depends only on the system configuration, enabling the OP to be evaluated under arbitrary channel correlation models. Overall, the main analytical results in this section are applicable to different correlation models, whereas the simple reference model is further investigated only to derive tractable approximations and enable efficient computation.

\subsection{Correlation Between Two Arbitrary Codewords}\label{Main Results-A}
We use $\boldsymbol{c}_u$ to represent the underlying random codeword {\em before normalization}. Assuming the codeword set $\mathcal{U}=\{\boldsymbol{c}_u, u=1,\ldots,U\}$, where each codeword is independently generated with elements following $\mathcal{CN}(0,\sigma_c^2)$, the correlation between two arbitrary codewords is defined by
\begin{equation}\label{eq:6}
\rho_{i,j} = \frac{|\boldsymbol{c}_i^{\mathrm{H}}\boldsymbol{c}_j|}{\|\boldsymbol{c}_i\|_2\|\boldsymbol{c}_j\|_2},~i\neq j,
\end{equation} 
where it should be emphasized that $\rho_{i,j}$ in \eqref{eq:6} is a realization-wise geometric similarity measure between two finite-length codewords, rather than the ensemble-level statistical correlation implied by the random codebook generation model.

It is worth noting that although the codewords are generated independently and therefore are uncorrelated in the statistical sense, the correlation defined in \eqref{eq:6} represents the normalized inner product of two specific realizations. Consequently, $\rho_{i,j}$ is generally non-zero for finite $M$, and its distribution and extreme values are of practical interest. The averaged and the maximum correlation are defined as
\begin{subequations}\label{eq:7}
\begin{align}
\bar{\rho}&=\frac{2}{U\left(U-1\right)}\sum_{i<j}\rho_{i,j},\label{eq:7a}\\
\rho_{\max}& = \max_{i\neq j}\rho_{i,j}.\label{eq:7b}
\end{align}
\end{subequations}

Different from prior results in which the correlation among users' signals was not considered, we argue that such correlation influences the inter-user interference and the calculation of each user's SINR, to be proven in \eqref{eq:SINR_final_robust} and \eqref{eq:MRC_SINR_Final}. We address this issue using the results from Theorem~\ref{Theorem 1} to Theorem~\ref{Theorem 4}.

\begin{theorem}\label{Theorem 1}
Let $\boldsymbol{c}_i$ and $\boldsymbol{c}_j$ be two\footnote{It means any distinct pair selected from the independently generated random codebook, rather than codewords with an arbitrary dependence structure.} distinct and independent codewords with independent and identically distributed (i.i.d.) entries distributed as $\mathcal{CN}(0,\sigma_c^2)$. Although each product term $c^{*}_{i,m}c_{j,m}$ is non-Gaussian, it has zero mean and finite variance. Hence, for moderate-to-large blocklength $M$, the inner product $\boldsymbol{c}_i^{\mathrm{H}}\boldsymbol{c}_j=\sum_{m=1}^{M}c^{*}_{i,m}c_{j,m}$ is well approximated by $\mathcal{CN}\!\left(0,M\sigma_c^4\right)$ according to the central limit theorem (CLT).
\end{theorem}

\begin{IEEEproof}
See Appendix~\ref{appen-a}.
\end{IEEEproof}
\begin{remark}
{\em The use of CLT in Theorem \ref{Theorem 1} is both theoretically and practically reasonable. Statistically, for random samples with identical distributions, the CLT is generally justified when the sample size is no less than $30$. According to \cite{CB_ML}, the channel coherence time is roughly $1/(4D_s)$, where $D_s$ is the maximal Doppler spread. For a 2 GHz carrier, it ranges from $1$ to $45~{\rm ms}$ for transmitter speeds in the range of $3$--$120~{\rm km/h}$. Combined with a typical outdoor sampling frequency of $100$--$500~{\rm kHz}$, this yields an FBL in the order of $10^2$--$10^4$, whereas port switching takes only $\mu s$ \cite[Fig.~11]{channel_scan}.}
\end{remark}

\begin{corollary}\label{Corollary 1}
Under CLT in Theorem~\ref{Theorem 1}, the amplitude $|\boldsymbol{c}_i^{\mathrm{H}}\boldsymbol{c}_j|$ is approximately Rayleigh distributed with PDF
\begin{equation}\label{eq:8}
p_{|\boldsymbol{c}_i^{\mathrm{H}}\boldsymbol{c}_j|}\left(r\right)=\frac{2r}{M\sigma_c^4}e^{-\frac{r^2}{M\sigma_c^4}},~r\ge0,
\end{equation}
by which, we have $\mathrm{E}\left[|\boldsymbol{c}_i^{\mathrm{H}}\boldsymbol{c}_j|\right]=\sigma_c^2\frac{\sqrt{\pi M}}{2}$.
\end{corollary}

\begin{theorem}\label{Theorem 2}
The squared $\ell_2$-norm of a codeword follows a Gamma distribution, i.e., $\|\boldsymbol{c}_i\|_2^2 \sim \mathcal{G}(M,\sigma_c^2)$, which characterizes the codeword energy, where $M$ denotes the codeword length and $\sigma_c^2$ is the variance of each codeword entry. Moreover, practically for $M\gg 1$, by the CLT, the distribution of $\|\boldsymbol{c}_i\|_2^2$ can be approximated as $\|\boldsymbol{c}_i\|_2^2 \sim \mathcal{N}\!\left(M\sigma_c^2,M\sigma_c^4\right)$.
\end{theorem}

\begin{IEEEproof}
See Appendix~\ref{appen-b}.
\end{IEEEproof}

\begin{corollary}\label{Corollary 2}
On average, the squared $\ell_2$-norm of the codeword satisfies $\mathrm{E}[\|\boldsymbol{c}_i\|_2^2] = M\sigma_c^2$. By the law of large numbers, for sufficiently large $M$, the squared norm concentrates around its mean, i.e., $\|\boldsymbol{c}_i\|_2^2 \approx M\sigma_c^2$, which implies that the $\ell_2$-norm of the codeword itself is approximately $\|\boldsymbol{c}_i\|_2 \approx \sigma_c\sqrt{M}$. Since it has been proven that $\boldsymbol{c}_i^{\mathrm{H}}\boldsymbol{c}_j \sim \mathcal{CN}(0, M\sigma_c^4)$, the normalized inner product is given by $\frac{\boldsymbol{c}_i^{\mathrm{H}}\boldsymbol{c}_j}{\|\boldsymbol{c}_i\|_2\|\boldsymbol{c}_j\|_2} \sim \mathcal{CN}\left(0, \frac{1}{M}\right)$.
\end{corollary}

\begin{theorem}\label{Theorem 3}
	The expected average correlation between arbitrary pairs of codewords in \eqref{eq:7a} can be approximated as
	\begin{equation}\label{eq:average_correlation}
		\bar{\rho}
		=\frac{2}{U(U-1)}\sum_{i<j}\rho_{i,j}
		\approx \sqrt{\frac{\pi}{4M}}.
	\end{equation}
\end{theorem}

\begin{IEEEproof}
From Corollary~\ref{Corollary 1} where $\mathrm{E}\left[|\boldsymbol{c}_i^{\mathrm{H}}\boldsymbol{c}_j|\right]=\sigma_c^2\frac{\sqrt{\pi M}}{2}$ and Corollary~\ref{Corollary 2} where $\|\boldsymbol{c}_i\|_2\approx \sigma_c\sqrt{M}$, since sample average converges into statistical expectation, we have $\bar{\rho}\rightarrow\mathrm{E}\left[\rho_{i,j}\right]=\mathrm{E}\left[\frac{|\boldsymbol{c}_i^{\mathrm{H}}\boldsymbol{c}_j|}{\|\boldsymbol{c}_i\|_2\|\boldsymbol{c}_j\|_2}\right]\approx \frac{\sigma_c^2\frac{\sqrt{\pi M}}{2}}{\left(\sigma_c\sqrt{M}\right)^2}=\sqrt{\frac{\pi}{4M}}$.
\end{IEEEproof}

\begin{theorem}\label{Theorem 4}
The average maximum correlation defined in \eqref{eq:7b} over all codeword pairs can be approximated as
\begin{equation}\label{eq:10}
\begin{aligned}
	&\bar{\rho}_{\max}=\mathrm{E}\left[\rho_{\max}\right]\approx\sqrt{\frac{\ln T}{M}}+\frac{\gamma}{2\sqrt{M\ln T}},\\
	&\mbox{and}~T=\frac{U(U-1)}{2},
\end{aligned}
\end{equation}
where $\gamma \approx 0.57721$ is the Euler-Mascheroni constant.
\end{theorem}

\begin{IEEEproof}
See Appendix~\ref{appen-c}.
\end{IEEEproof}
\begin{figure}[t!]
	\centering
	\includegraphics[width=\columnwidth]{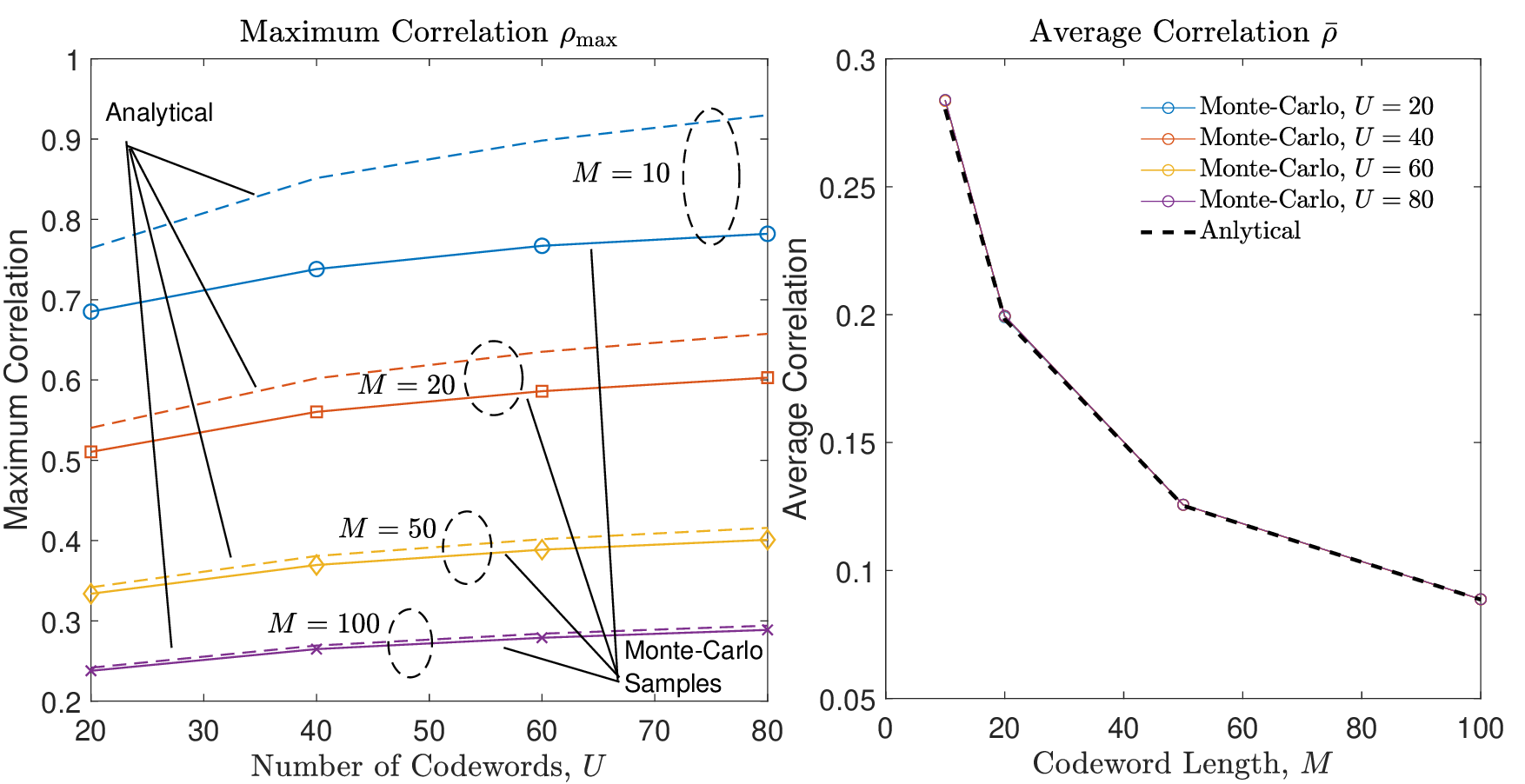}
	\caption{Illustration of analytical codeword correlation (dashed lines) under FBL with $2,000$ Monte-Carlo (solid lines), codeword length $M\in\{10,20,50,100\}$ and number of codewords $U\in \{20,40,60,80\}$. On the left: maximum correlation $\rho_{\max}$ by \eqref{eq:10} of Theorem~\ref{Theorem 4}. On the right: average correlation $\bar{\rho}$ by \eqref{eq:average_correlation} of Theorem~\ref{Theorem 3}.}\label{fig:codeword_correlation}
\end{figure}
In Fig.~\ref{fig:codeword_correlation}, the analytical approximations of $\rho_{\max}$ and $\bar{\rho}$ are presented for codeword lengths $M \in \{10,20,50,100\}$ and numbers of codewords $U \in \{20,40,60,80\}$. As expected, $\bar{\rho}$ is independent of $U$, and the estimation accuracy improves as the codeword length increases. For the maximum correlation, the analytical prediction agrees closely with the empirical results when $M$ is on the order of $10^2$. Interestingly, this trend is further supported by the results in \cite[Fig.~5]{codeword_design1} obtained using {\em practical Zadoff-Chu codewords, where the empirical CDF closely follows the theoretical bounds and more than $95\%$ of the correlation values remain below the level predicted by Theorem~\ref{Theorem 4} when the codeword length is on the order of $10^2$.} More simulations on codeword correlation are illustrated in Sec.~\ref{sim_code}.

\subsection{BLER Upper-Bound for FBL-FAS}\label{Main Results-B}
Here, we upper-bound the BLER of FAS in the FBL regime, with results applicable to all the existing correlation models. 
To start with, we rewrite \eqref{eq:5} into a more compact form:
\begin{equation}\label{eq:11}
\boldsymbol{y}=\sum_{u=1}^{U}g_{u,k}\boldsymbol{x}_u+\boldsymbol{\eta}=\boldsymbol{X}\boldsymbol{g}+\boldsymbol{\eta},
\end{equation}
with codeword (column of $\boldsymbol{X}$) $\boldsymbol{x}_u\in \mathcal{U}$. We assume an idealized port selection that chooses the strongest port. Notably, in quasi-static fading, the channel scanning or port selection can be carried out prior to payload transmission. The impact of practical switching/estimation overhead is an important topic but beyond the scope of this work \cite[Fig.~11]{channel_scan}. Let $g_{u,\mathrm{FAS}}$ denote the channel response after port selection, given by
\begin{equation}\label{eq:12}
|g_{u,\mathrm{FAS}}| = \max\left\{|g_{u,1}|,|g_{u,2}|,\ldots,|g_{u,N}|\right\}.
\end{equation}
Although each user performs port selection independently and thus obtains a possibly different realization $g_{u,\mathrm{FAS}}$, the random variables $|g_{u,\mathrm{FAS}}|$ are i.i.d.~across $u$ under independent user channels. For notational simplicity, we denote a generic post-selection gain by $|g_{\mathrm{FAS}}|$ when only its distribution is involved.

With FBL, each user transmits one codeword within a single block, and an erroneous detection of any user's codeword results in a block error.  At the receiver, we employ a block-wise ML detector, which jointly detects the transmitted codewords by minimizing the Euclidean distance between the received signal and all possible superpositions of candidate codewords. 

Specifically, consider an error event in which a subset of users, denoted by
$\mathcal{U}'\subseteq\mathcal{U}$ with cardinality
$\lvert\mathcal{U}'\rvert=U'$, is detected incorrectly. The remaining users
in the complementary set, $\mathcal{U}\setminus\mathcal{U}'$ with cardinality
$\lvert\mathcal{U}\setminus\mathcal{U}'\rvert=U-U'$, are correctly detected.
An erroneous hypothesis is formed by replacing the actually transmitted
codewords of the users in $\mathcal{U}'$ with an incorrect codeword collection
$\boldsymbol{X}'$, while the remaining users in
$\mathcal{U}\setminus\mathcal{U}'$ are correctly detected. The error event
$\mathcal{E}_{\mathcal{U}'}$ occurs when the ML detector favors this erroneous
hypothesis over the true one, i.e., when the received signal is closer to the
erroneous hypothesis than to the true hypothesis (in the Euclidean-distance
sense). In particular, the event $\mathcal{E}_{\mathcal{U}'}$ denoting $U'$
erroneous detections occurs if
\begin{subequations}\label{eq:13}
	\begin{align}
		\mathcal{E}_{{\mathcal{U}}'}&=\begin{Bmatrix}
			\|\boldsymbol{y}-\underbrace{{\boldsymbol{X}}'{\boldsymbol{g}}'}_{
				\text{${U}'$ error}}-\underbrace{\boldsymbol{X}_{\mathcal{U}\setminus{\mathcal{U}}'}\boldsymbol{g}_{\mathcal{U}\setminus{\mathcal{U}}'}}_{\text{$U-{U}'$ correctly detected}}\|^2_2 \\
			< \|\boldsymbol{y}-\boldsymbol{X}\boldsymbol{g}\|^2_2		
		\end{Bmatrix}, \label{eq:13a}\\
		&=\left\{\|\underbrace{\left(\boldsymbol{X}_{\mathcal{U}'}-{\boldsymbol{X}}'\right){\boldsymbol{g}}'}_{\boldsymbol{\eta}'}+\boldsymbol{\eta}\|^2_2 < \|\boldsymbol{\eta}\|^2_2 \right\}, \label{eq:13b}
	\end{align}
\end{subequations}
where $\boldsymbol{X}_{\mathcal{U}'}$ denotes the actually transmitted codewords of the users belonging to the error set $\mathcal{U}'$ and $\boldsymbol{g}'$ denotes the corresponding channel coefficient vector. Here, the left-hand side of \eqref{eq:13b} is obtained by substituting the true received signal $\boldsymbol{y}=\boldsymbol{X}_{\mathcal{U}'}\mathbf{g}'+\boldsymbol{X}_{\mathcal{U}\setminus\mathcal{U}'}\boldsymbol{g}_{\mathcal{U}\setminus\mathcal{U}'}+\boldsymbol{\eta}$ into the left-hand side of \eqref{eq:13a}. Considering all possible combinations of $\mathcal{U}'$, the universal set of error events is denoted by $\mathcal{W}=\{\mathcal{U}'_i,\ i=1,2,\ldots\}$.

Furthermore, we denote $\boldsymbol{\eta}'=\left(\boldsymbol{X}_{\mathcal{U}'}-{\boldsymbol{X}}'\right){\boldsymbol{g}}'$ in the sequel. Meanwhile, there are $|\mathcal{W}|=\binom{U}{{U}'}^2 $ possible combinations for ${\mathcal{U}'}$ \cite[(13)]{chernoff1}, \cite[(12)]{chernoff2}. Thus, the BLER for FBL-FAS can be calculated as
\begin{equation}\label{eq:14}
\mathrm{BLER}=P\left(\bigcup_{ {\mathcal{U}}' }\mathcal{E}_{{\mathcal{U}}'}\right)\le |\mathcal{W}|P\left(\mathcal{E}_{{\mathcal{U}}'}\right).
\end{equation}

\begin{remark}
\textit{Note that the second inequality in \eqref{eq:14} relies on the fact that all error events $\{\mathcal{E}_{\mathcal{U}'}\}$ with the same cardinality $|\mathcal{U}'|=U'$ are equiprobable. This follows from the symmetry of the system model that all users employ i.i.d.~Gaussian codewords which are detected using a permutation-invariant ML decision rule. Therefore, the probability of an erroneous detection event depends only on the number of incorrectly detected users $U'$, rather than on their specific indices, which justifies the multiplication by $|\mathcal{W}|$.}
\end{remark}

\begin{theorem}\label{Theorem 5}
The closed-form BLER conditioned on the instantaneous $|g_{\mathrm{FAS}}|$ is upper-bounded by
\begin{equation}\label{eq:15}
P\left(\mathcal{E}_{{\mathcal{U}}'}\Big\vert |g_{\mathrm{FAS}}|\right)\le \sum_{{U}'=0}^{U}\frac{{U}'}{U}e^{{L}'-M\log\left(1+\frac{0.25M\sigma_{\eta'}^2}{\sigma_{\eta}^2}\right)},
\end{equation}
in which ${L}'=\log\binom{U}{{U}'}^2=\sum_{i=1}^{{U}'-1}2\log\frac{U-i}{{U'-i}}$ and $\sigma^2_{\eta'}=2{U}'\sigma_c^2|g_{\mathrm{FAS}}|^2$
\end{theorem}

\begin{IEEEproof}
See Appendix~\ref{appen-d}.
\end{IEEEproof}

\begin{figure}[]
\centering
\includegraphics[width=\columnwidth]{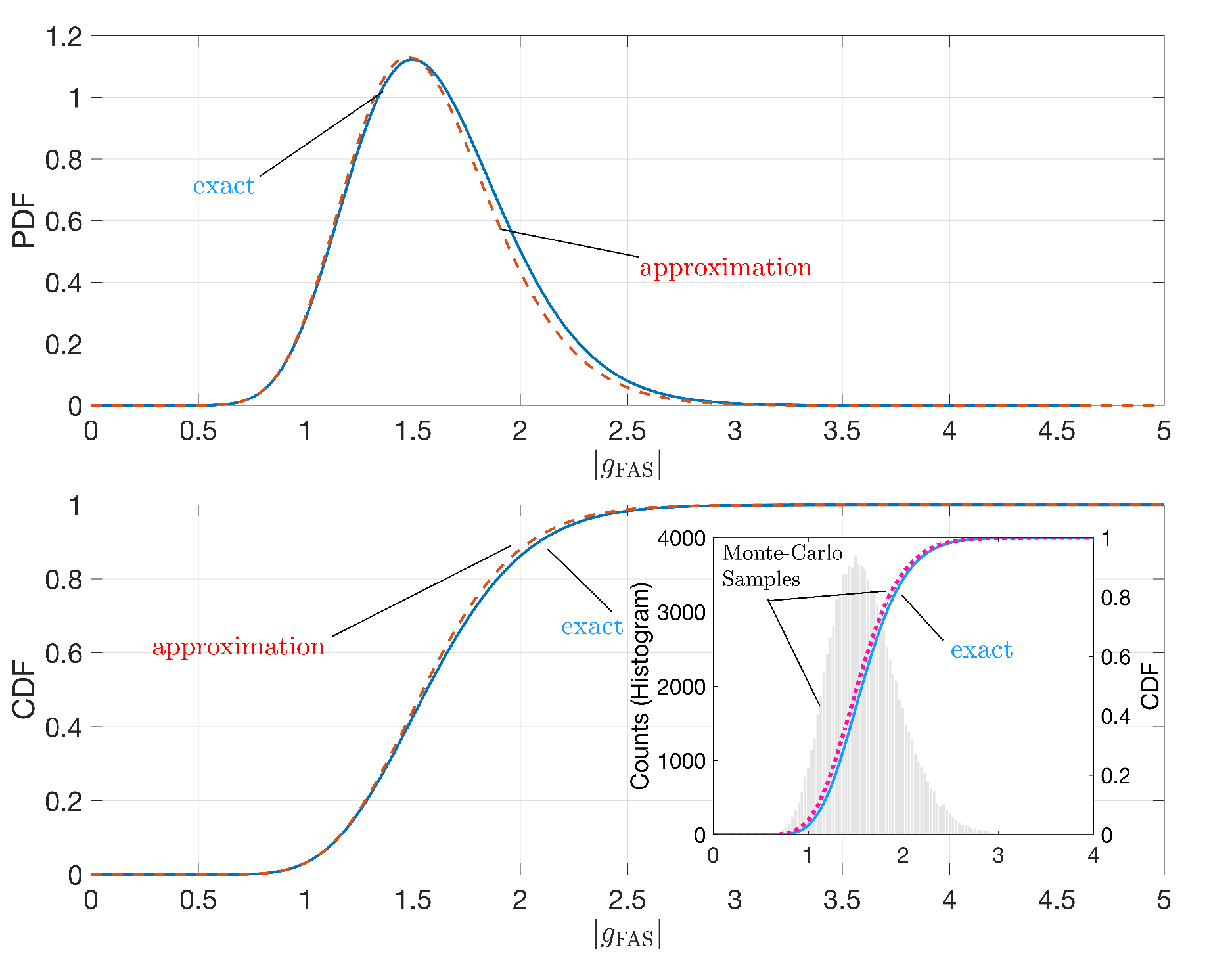}
\caption{Illustration of the simplified CDF and PDF of $|g_{\mathrm{FAS}}|$ in \eqref{simplifed_pdf_cdf} of Theorem~\ref{Theorem 7} with $10^{5}$ Monte-Carlo samples, $\sigma=1$, number of available ports $N=10$ and antenna length $W=0.5$.}\label{fig:PDF_CDF_Approximation}
\end{figure}

\begin{figure*}[]
\begin{multline}\label{eq:16}
	\begin{aligned}
		&C_{|g_{\mathrm{FAS}}|}(r)= \int_0^{\frac{r^2}{\sigma^2}} e^{-t} \prod_{k=2}^N \left[ 1 - Q_1\left( \sqrt{\frac{2 \mu_k^2}{1 - \mu_k^2}} \sqrt{t}, \sqrt{\frac{2}{\sigma^2 (1 - \mu_k^2)}} r \right) \right] \mathrm{d}t,\\
&f_{|g_{\mathrm{FAS}}|}(r) =\\
&\frac{2r}{\sigma^2} e^{-\frac{r^2}{\sigma^2}} \prod_{k=2}^N \left[ 1 - Q_1\left( \sqrt{\frac{2 \mu_k^2}{1 - \mu_k^2}} \frac{r}{\sigma}, \sqrt{\frac{2}{1 - \mu_k^2}} \frac{r}{\sigma} \right) \right]+
\sum_{i=2}^N \int_0^{\frac{r^2}{\sigma^2}} e^{-t} \left[ \prod_{k=2, k \neq i}^N \left( 1 - Q_1\left( \sqrt{\frac{2 \mu_k^2}{1 - \mu_k^2}} \sqrt{t}, \sqrt{\frac{2}{\sigma^2 (1 - \mu_k^2)}} r \right) \right) \right] \\
&\times\frac{2r}{1 - \mu_i^2} \frac{1}{\sigma^2} \left( e^{ -\frac{\frac{2 \mu_i^2 t}{1 - \mu_i^2} + \frac{2 r^2}{\sigma^2 (1 - \mu_i^2)}}{2} }\right) I_0\left( \sqrt{\frac{2 \mu_i^2 t}{1 - \mu_i^2}}\sqrt{\frac{2}{1 - \mu_i^2}} \frac{r}{\sigma} \right)\mathrm{d}t
	\end{aligned}
\end{multline}
\hrulefill
\end{figure*}

\begin{figure*}[]
\begin{equation}\label{simplifed_pdf_cdf}
\begin{aligned}
&C_{|g_{\mathrm{FAS}}|}(r)\approx
(1-e^{-\frac{r^2}{\sigma^2}})\prod_{k=2}^N\left[1-Q_1(\sqrt{\frac{2\mu_k^2}{1-\mu_k^2}}\sqrt{\bar{t}},\sqrt{\frac{2}{\sigma^2(1-\mu_k^2)}}r)\right],\\
&f_{|g_{\mathrm{FAS}}|}(r) \approx\\ &\frac{2r}{\sigma^2} e^{-\frac{r^2}{\sigma^2}} \prod_{k=2}^N \left[ 1 - Q_1\left( \sqrt{\frac{2 \mu_k^2}{1 - \mu_k^2}} \frac{r}{\sigma}, \sqrt{\frac{2}{1 - \mu_k^2}} \frac{r}{\sigma} \right) \right]
+\left(1-e^{-\frac{r^2}{\sigma^2}}\right)\sum_{i=2}^N \left[ \prod_{k=2, k \neq i}^N \left( 1 - Q_1\left( \sqrt{\frac{2 \mu_k^2}{1 - \mu_k^2}} \sqrt{\bar{t}}, \sqrt{\frac{2}{\sigma^2 (1 - \mu_k^2)}} r \right) \right) \right] \\ &\times\frac{2r}{1 - \mu_i^2} \frac{1}{\sigma^2} \left( e^{ -\frac{\frac{2 \mu_i^2 \bar{t}}{1 - \mu_i^2} + \frac{2 r^2}{\sigma^2 (1 - \mu_i^2)}}{2} }\right) I_0\left( \sqrt{\frac{2 \mu_i^2 \bar{t}}{1 - \mu_i^2}}\sqrt{\frac{2}{1 - \mu_i^2}} \frac{r}{\sigma} \right),~\mbox{where }\bar{t}=\frac{1-\left(\frac{r^2}{\sigma^2}+1\right)e^{-\frac{r^2}{\sigma^2}}}{1-e^{-\frac{r^2}{\sigma^2}}},
\end{aligned}
\end{equation}
\hrulefill
\end{figure*}

Note that \eqref{eq:15} in Theorem~\ref{Theorem 5} enables \textit{instantaneous BLER calculation for all channel modelings in Section \ref{sec:system model}}, since it only requires $|g_{\mathrm{FAS}}|$ from randomized realizations. If the PDF of any channel modeling is available, the statistically averaged BLER can be obtained in Theorem~\ref{Theorem 8}. Moreover, we derive and simplify the PDF and CDF of a simple reference correlation model by Theorem~\ref{Theorem 6} and Theorem~\ref{Theorem 7}, respectively.

\begin{theorem}\label{Theorem 6}
The exact CDF and PDF under the simplified reference modeling of $|g_{\mathrm{FAS}}|$ is given by \eqref{eq:16} (see top of next page).
\end{theorem}

\begin{IEEEproof}
See Appendix~\ref{appen-e}.
\end{IEEEproof}

\begin{theorem}\label{Theorem 7}
Based on {\em MVTI} and {\em Taylor expansion}, the approximated CDF and PDF of $|g_{\mathrm{FAS}}|$ under the simplified reference modeling is given by \eqref{simplifed_pdf_cdf} (see next page).
\end{theorem}

\begin{IEEEproof}
See Appendix~\ref{appen-f}.
\end{IEEEproof}

Fig.~\ref{fig:PDF_CDF_Approximation} illustrates and compares the CDFs (exact, approximated and Monte-Carlo) and PDFs (exact and approximated) with setups $\sigma=1$, $N=10$ and $W=0.5$. The simplifed CDF and PDF indicate marginal estimation loss compared with the exact ones but particularly, the simplifed expressions {\em are free from the complicated product-integral calculations of \eqref{eq:16}}.

\begin{theorem}\label{Theorem 8}
The BLER for FBL-FAS is upper-bounded by
\begin{equation}\label{eq:17}
\begin{aligned}
\mathrm{BLER}&=\int f_{|g_{\mathrm{FAS}}|}(r) P\left(\mathcal{E}_{{\mathcal{U}}'}\Big\vert |g_{\mathrm{FAS}}|\right) \mathrm{d}r\\ 
&\le\sum_{{U}'=0}^{U}\int_{r=0}^{+\infty}f_{|g_{\mathrm{FAS}}|}(r)\times \frac{{U}'}{U} e^{{L}'-M\log\left(1+\frac{0.25M\sigma_{\eta'}^2}{\sigma_{\eta}^2}\right)} \mathrm{d}r,
\end{aligned}
\end{equation}
where $\sigma^2_{\eta'}=2{U}'\sigma_c^2r^2$, and ${L}'$ is already given in Theorem \ref{Theorem 5}.
\end{theorem}

\begin{remark}
\emph{Theorem~\ref{Theorem 8} demonstrates the generality of the conditional BLER expression in Theorem~\ref{Theorem 5}, which is further illustrated in Fig.~\ref{fig:BLER_models}. For any correlation model characterized by the PDF $f_{|g_{\mathrm{FAS}}|}(r)$, it provides the corresponding average BLER by averaging over the post-selection channel distribution. In particular, by using the distributions in Theorem~\ref{Theorem 6} and Theorem~\ref{Theorem 7}, the average BLER under the simplified reference correlation model can be readily obtained.}
\end{remark}

\subsection{Redefined OP for FBL-FAS}\label{sec:OP_FAS}
From \cite[Theorem~3]{FBL_MIMO}, the OP under FBL should be defined as the probability that the instantaneous mutual information is smaller than the desirable code rate, i.e., 
\begin{equation}\label{eq.OP_def}
p^{}_{\mathrm{out}}(\gamma_{\mathrm{th}})=P\left\{\log_2\left(1+\Gamma\right)\le R_c\right\},
\end{equation}
where $\Gamma$ is the SINR per channel use and $R_c$ is the code rate.

\begin{theorem}\label{Theorem 9}
The OP for FBL-FAS can be calculated by
\begin{align}
p^{\mathrm{FBL\text{-}FAS}}_{\mathrm{out}}(\gamma^{\mathrm{FAS}}_{\mathrm{th}})&\triangleq P\left\{\log_2\left(1+\Gamma_{\mathrm{FAS}}\right)\le R_c\right\}\notag\\
&= P\left\{|g_{\mathrm{FAS}}|\le \gamma^{\mathrm{FAS}}_{\mathrm{th}}\right\}\label{eq:OP_function}
\end{align}
with the outage threshold
\begin{equation}
\gamma^{\mathrm{FAS}}_{\mathrm{th}}=\sqrt{ \left(2^{R_c}-1\right)\left(\left(U-1\right)\left(\sigma^2+\left(U-2\right)\frac{\pi}{4}\sigma^2\bar{\rho}\right)+\sigma^2_{\eta}\right)},
\end{equation} 
which is a constant, {\em irrelevant to correlation models.}
\end{theorem}
\begin{IEEEproof}
See Appendix~\ref{appen-h}.
\end{IEEEproof}

\begin{remark}
{\em If the simple reference correlation model is used, then the corresponding OP can be calculated by
\begin{multline}
P\left\{|g_{\mathrm{FAS}}|\le \gamma^{\mathrm{FAS}}_{\mathrm{th}}\right\}=\int_0^{\frac{\left(\gamma^{\mathrm{FAS}}_{\mathrm{th}}\right)^2}{\sigma^2}} e^{-t}\\
\times \prod_{k=2}^N \left[ 1 - Q_1\left( \sqrt{\frac{2 \mu_k^2}{1 - \mu_k^2}} \sqrt{t}, \sqrt{\frac{2}{\sigma^2 (1 - \mu_k^2)}} \gamma^{\mathrm{FAS}}_{\mathrm{th}} \right) \right] \mathrm{d}t.
\end{multline}}
\end{remark}

\section{Benchmark: Conventional $L$-FPA System}\label{sec:benchmark_metric}
Here, the BLER and OP expressions of the conventional $L$-FPA system are approximated. Note that this system requires $L$ radio-frequency (RF) chains at each user but the considered FAS needs only one RF chain.

\begin{remark}\label{Remark 1}
{\em In accordance with \cite[(1)]{intro_FBL1}, the BLER using random coding is approximated as
\begin{equation}\label{eq:minimum_FBL_BLER}
\epsilon_L\approx Q\left(\frac{C-R_c}{\sqrt{V_{\rm dis}/M}}\right),
\end{equation}
where $C=\frac{1}{2}\log_2\left(1+\Gamma_{\mathrm{MRC}}\right)$ is the averaged channel capacity, $V_{\rm dis}=\frac{\Gamma_{\mathrm{MRC}}}{2}\frac{\Gamma_{\mathrm{MRC}}+2}{(\Gamma_{\mathrm{MRC}}+1)^2}\log_2^2(e)$ is the channel dispersion and $R_c=\frac{\log_2U}{M}$ denotes the code rate \cite[Theorem 1]{Sparse_code}, \cite{SC2}. For $L$-FPA systems using MRC, the SINR, $\Gamma_{\mathrm{MRC}}$, equals to \eqref{eq:MRC_SINR_Final}. A closed-from BLER for conventional systems in \eqref{eq:minimum_FBL_BLER} offers an accurate performance prediction when the blocklength is relatively large in the order of $10^2$.}
\end{remark}

Assuming independence among users' channels, the MRC SINR, $\Gamma_{\mathrm{MRC}}$, can be calculated as \cite{SINR_MRC_2}
\begin{equation}\label{eq:upper_SINR}
\begin{aligned}
\Gamma_{\mathrm{MRC}} &=\mathrm{E}\left\{\frac{\frac{1}{M}\|\boldsymbol{g}_i\|^4_2}{\frac{1}{M}\sum_{u\neq i}|\boldsymbol{g}^{\mathrm{H}}_i\boldsymbol{g}_u|^2+\frac{\sigma_{\eta}^2}{M}\|\boldsymbol{g}_i\|^2_2}\right\}\\
&\le
\frac{4L^3\sigma^2}{\pi\left(U-1\right)\sigma^2+4L^2\sigma^2_{\eta}}.
\end{aligned}
\end{equation}

\begin{theorem}\label{Theorem 10}
The MRC-outage probability for conventional systems with $L$ FPAs, can be calculated as
\begin{equation}\label{eq:OP_function_MRC}
\begin{aligned}
p^{\mathrm{MRC}}_{\mathrm{out}}(\gamma^{\mathrm{MRC}}_{\mathrm{th}}) &= P\left\{ \log_2(1 + \Gamma_{\mathrm{MRC}}) \le R_c \right\}\\
& = P\left\{ \|\boldsymbol{g}_i\|^2 \le \gamma^{\mathrm{MRC}}_{\mathrm{th}} \right\}\\
&= 1 - \exp\left( - \frac{\gamma^{\mathrm{MRC}}_{\mathrm{th}}}{\sigma^2} \right) \sum_{k=0}^{L-1} \frac{1}{k!} \left( \frac{\gamma^{\mathrm{MRC}}_{\mathrm{th}}}{\sigma^2} \right)^k,
\end{aligned}
\end{equation}
with the outage threshold 
\begin{equation}
\gamma^{\mathrm{MRC}}_{\mathrm{th}}=\left (\left(U-1 \right)\left( \sigma^2  + \sigma^2 (U-2)\frac{\pi}{4}\bar{\rho} \right) + \sigma^2_{\eta}\right)\left(2^{R_c}-1\right),
\end{equation} 
which is a constant under fixed system configurations.
\end{theorem}

\begin{IEEEproof}
See Appendix~\ref{appen-i}.
\end{IEEEproof}

\section{Numerical Results}\label{sec: numerical results}
This section compares the BLER and OP performance of the single-antenna FBL-FAS with that of a conventional $L$-FPA MRC system. All parameter settings are provided in the corresponding figure captions. Several correlation models in Sec.~\ref{sec.models} are considered to illustrate the universality of the results developed in this work . For the OP evaluation under the fully correlated model, the PDF expression derived using the block-correlation approximation in \cite{block1} is given in \cite[(8)]{block_FBL_FAS}. Apparently, FAS operates only on one RF chain while the MRC system employs $L$ RF chains so the latter serves as a performance upper bound here and our interest is to find out whether FAS still achieves competitive performance, compared to a multi-FPA system.

\subsection{Codeword-Correlation Analysis}\label{sim_code}
\begin{figure}[t!]
	\centering
	\includegraphics[width=\columnwidth]{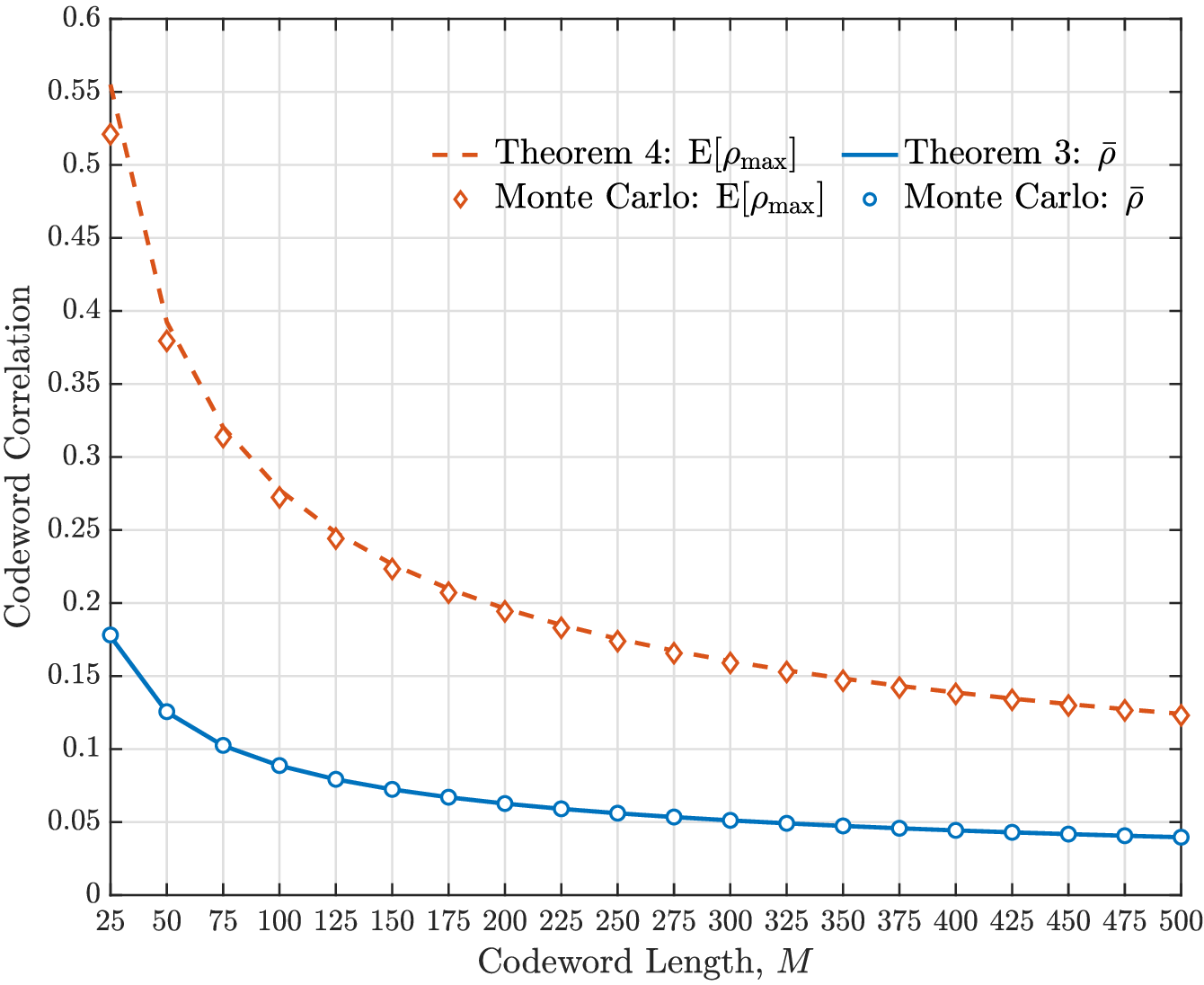}
	\caption{Average and maximum codeword correlations against the codeword length $M$ with $U=50$ and $M\in\{25,50,\ldots,500\}$. The curves represent the analytical results in Theorems~\ref{Theorem 3} and~\ref{Theorem 4}, while the markers represent the corresponding Monte Carlo results.}
	\label{fig:codeword_correlation_M}
\end{figure}

{\em Impact of Codeword Length}---Fig.~\ref{fig:codeword_correlation_M} compares the analytical and Monte Carlo results for the average correlation $\bar{\rho}$ and the average maximum correlation $\mathrm{E}[\rho_{\max}]$ under a fixed number of codewords $U=50$. Each codeword is independently generated with entries following $\mathcal{CN}(0,1)$. The Monte Carlo results closely follow the analytical expressions in Theorems~\ref{Theorem 3} and~\ref{Theorem 4}. In particular, Theorem~\ref{Theorem 3} accurately characterizes the average correlation over the entire considered blocklength range. The approximation in Theorem~\ref{Theorem 4} is slightly more pessimistic at the short blocklength (echoed by Fig.~\ref{fig:codeword_correlation}), e.g., around $M=25$, owing to the finite-$M$ approximations used in the norm concentration and extreme-value analysis, while the agreement becomes increasingly tight as $M$ grows. Both correlation measures decrease monotonically with $M$, as higher-dimensional Gaussian codewords become increasingly orthogonal. For fixed $U$, the number of codeword pairs remains unchanged, and both $\bar{\rho}$ and $\mathrm{E}[\rho_{\max}]$ exhibit an $M^{-1/2}$ decay.

\subsection{BLER Analysis}
\begin{figure}[]
\centering
\includegraphics[width=\columnwidth]{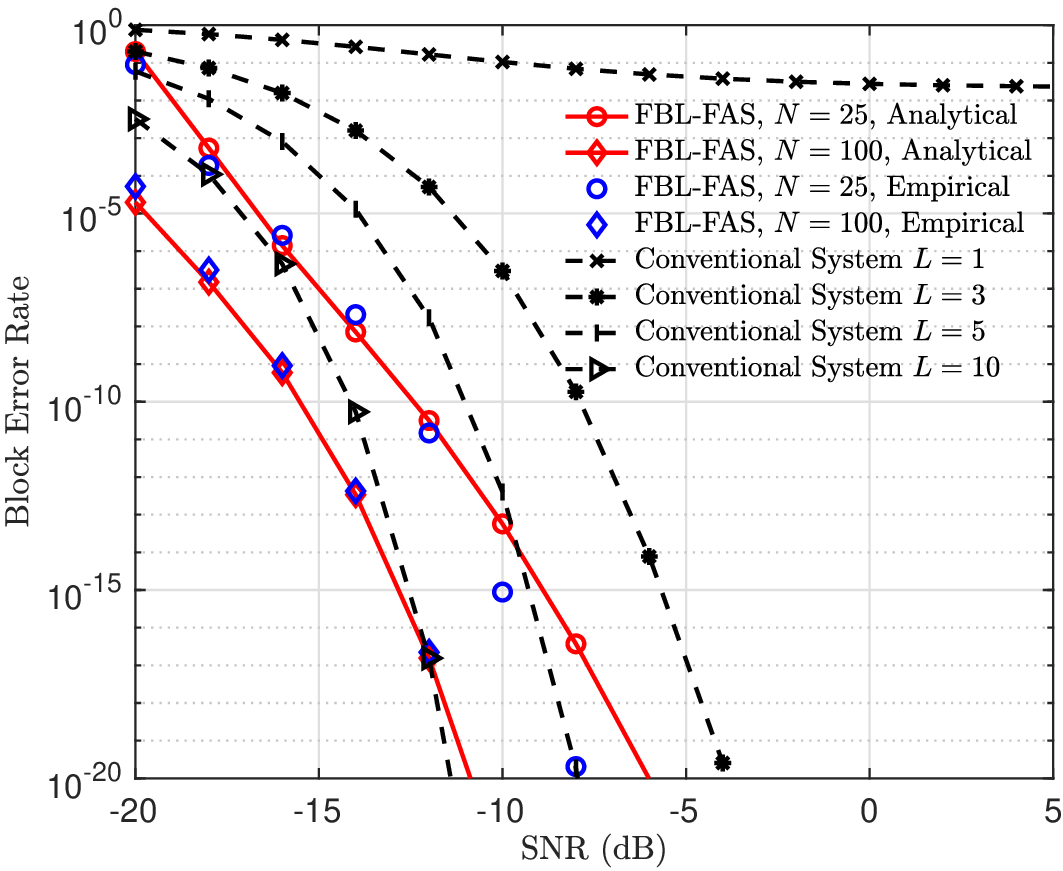}
\caption{BLER against SNR (dB) with $U=20, M=400, W=2, N\in\{25,100\}, L\in\{1,3,5,10\}$.}\label{fig:BLER_SNR}
\end{figure}

{\em Impact of System Parameters}---Fig.~\ref{fig:BLER_SNR} reveals that the conventional single-FPA system suffers from an error floor due to inter-user interference. In contrast, FAS yields substantial gains. Notably, with $N=100$, FAS achieves a $10^{5}$-fold BLER reduction compared to $N=25$ and significantly outperforms the $5$-FPA MRC system.

\begin{figure}[]
\centering
\includegraphics[width=\columnwidth]{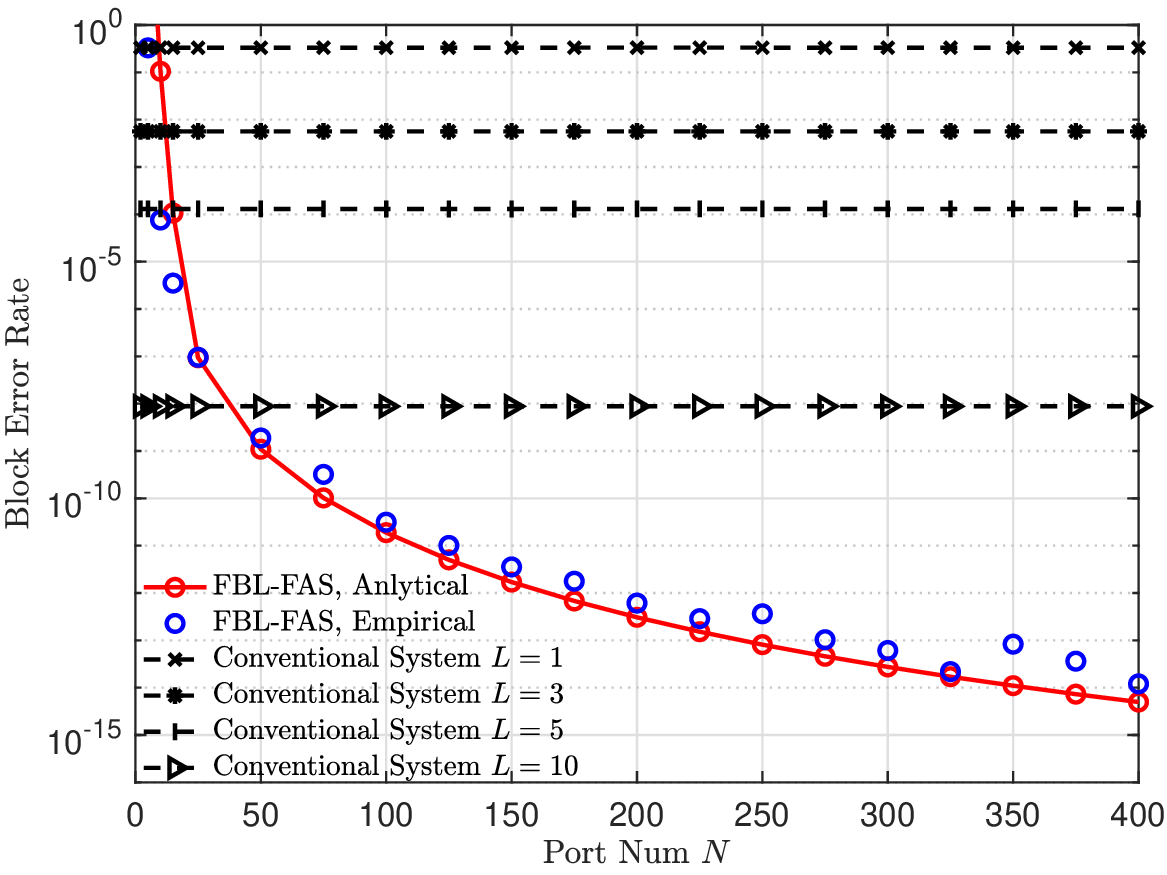}
\caption{BLER against port count $N$ with $U=20, M=400, W=2, \mathrm{SNR}=-15$ dB, $L\in\{1,3,5,10\}$.}\label{fig:BLER_N}
\end{figure}

\begin{figure}[]
\centering
\includegraphics[width=\columnwidth]{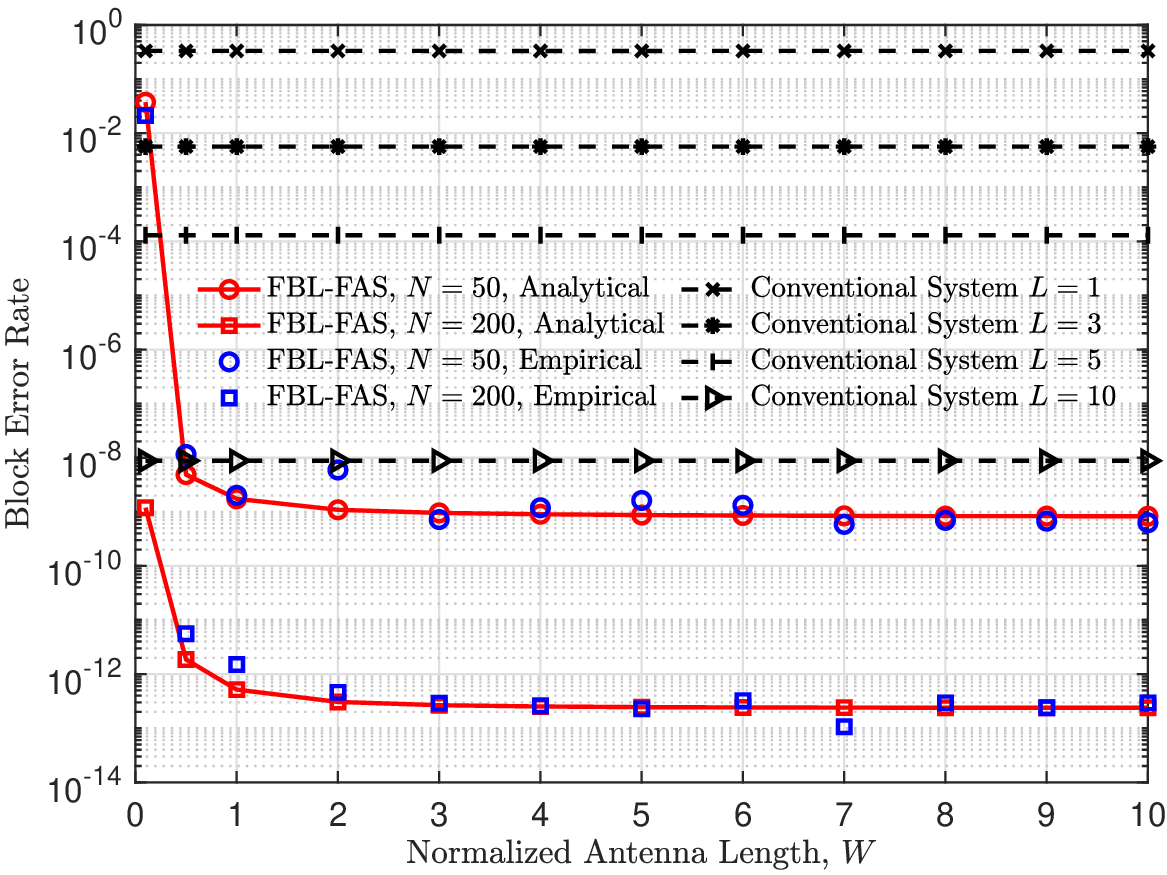}
\caption{BLER against aperture length $W$ with $M=400, \mathrm{SNR}=-15$ dB, $N\in\{50,200\}, L\in\{1,3,5,10\}$.}\label{fig:BLER_W}
\end{figure}

Fig.~\ref{fig:BLER_N} demonstrates the impact of port count $N$. As $N$ increases, spatial diversity improves significantly. Specifically, FAS with $N>50$ yields lower BLER than a $10$-FPA system. The effect of array size $W$ is examined in Fig.~\ref{fig:BLER_W}. With $N=50$ and $W=0.5$, FAS rivals the $L=10$ conventional system. Increasing $N$ further improves BLER by exploiting richer spatial diversity. Generally, when the total number of ports is fixed, increasing the normalized aperture enlarges the inter-port spacing. In the small-aperture regime, the ports are densely packed and strongly correlated. Hence, aperture expansion reduces the correlation and provides additional spatial degrees of freedom, leading to a noticeable performance improvement. However, once the port spacing becomes sufficiently large, the channel gains across the ports are already weakly correlated, and further aperture expansion yields little additional diversity gain. Consequently, the performance gradually saturates and approaches a floor as the normalized aperture continues to increase. Similarly, Fig.~\ref{fig:BLER_M} indicates that while the single-FPA system fails, FAS ($N=250$) achieves performance comparable to $L=10$ MRC with a blocklength of $M=160$.

\begin{figure}[]
\centering
\includegraphics[width=\columnwidth]{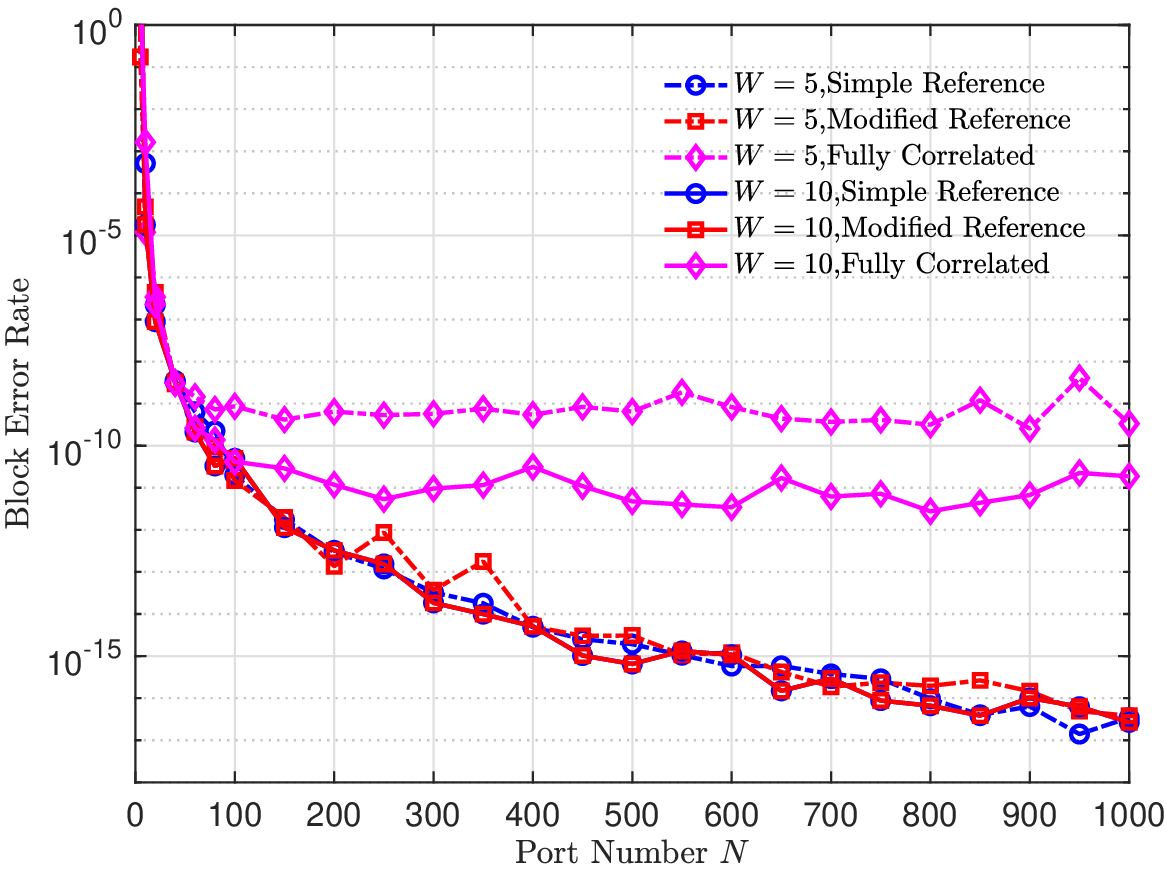}
\caption{BLER under different correlation models against $N$ with $M=400, U=20, W\in\{5,10\}, \mathrm{SNR}=-15$ dB.}\label{fig:BLER_models}
\end{figure}

\begin{figure}[]
	\centering
	\includegraphics[width=\columnwidth]{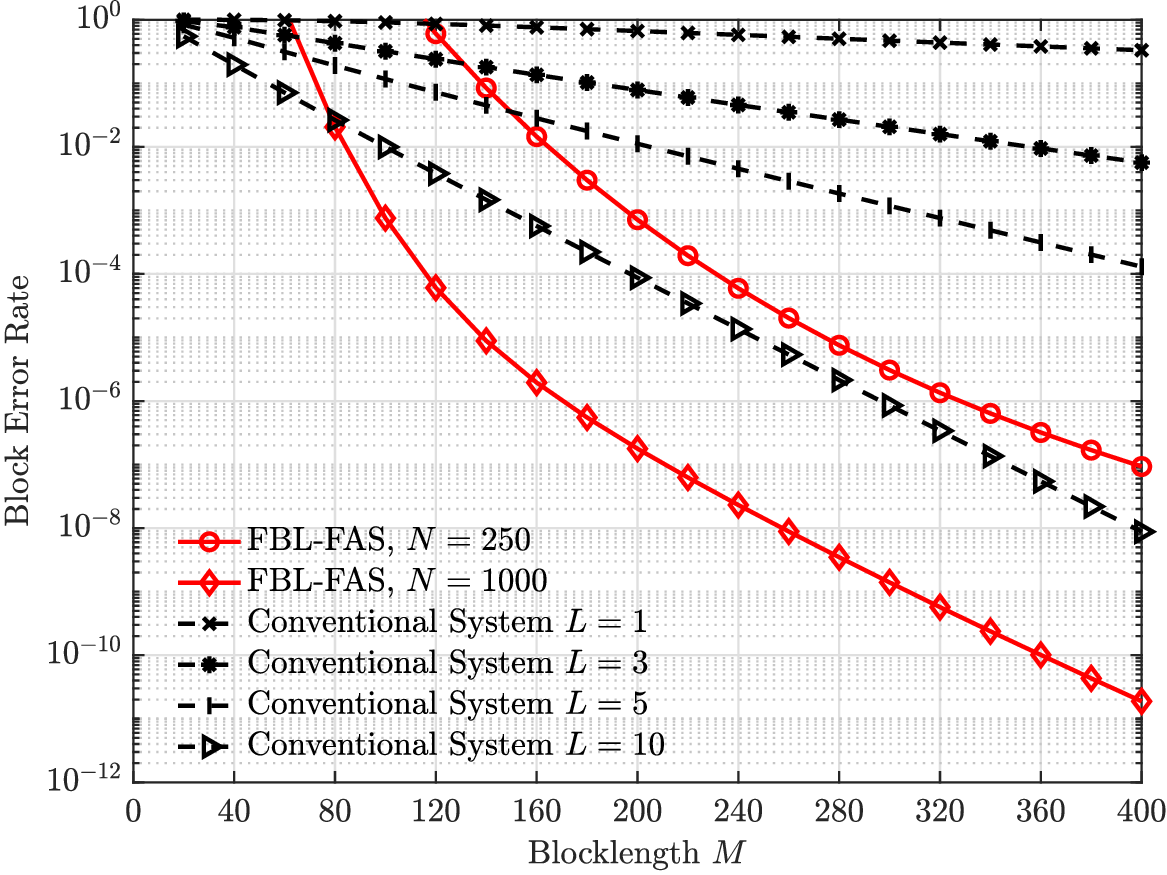}
	\caption{BLER against blocklength $M$ with $W=2, U=20, N\in\{250,1000\}, \mathrm{SNR}=-15$ dB, $L\in\{1,3,5,10\}$.}\label{fig:BLER_M}
\end{figure}

{\em Impact of Correlation Models}---Fig.~\ref{fig:BLER_models} validates the proposed bounds across different correlation models. For $N<100$, all models yield BLER around similar levels as correlation impact is negligible. The simple and modified reference models yield more optimistic BLER predictions because they do not fully capture the pairwise correlation among arbitrary ports \cite{fas_tutorial}, thereby overestimating the effective spatial diversity as $N$ increases. In contrast, the fully correlated model \cite{block1,block_FBL_FAS} in \eqref{eq:full} preserves the complete covariance structure and therefore exhibits a realistic performance floor when additional ports provide little new spatial diversity. Increasing $W$ reduces the correlation and lowers this floor. {\em More importantly, the proposed BLER framework is independent of any specific correlation model or closed-form channel PDF, and can be evaluated directly from analytically or empirically generated channel realizations, demonstrating its universal applicability to general FAS channel models.}

\subsection{OP Analysis}
\begin{figure}[]
	\centering
	\includegraphics[width=\columnwidth]{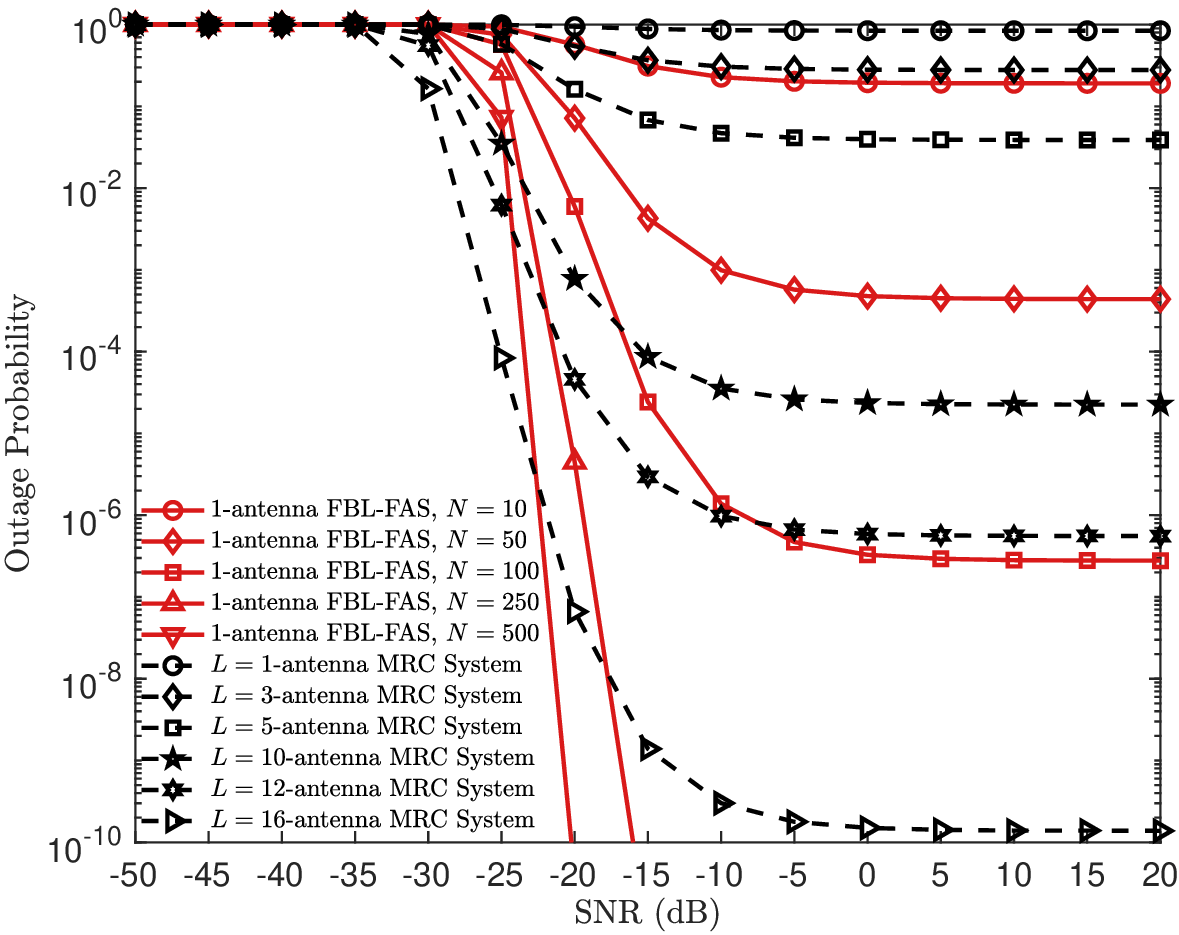}
	\caption{OP against SNR with $U=60, M=400, W=2, N\in \{10,50,100,250,500\}, L\in \{1,3,5,10,12,16\}$.}\label{fig:OP_SNR}
\end{figure}

Fig.~\ref{fig:OP_SNR} illustrates the OP performance. An error floor is seen for all curves due to the interference-limited SINR structure in \eqref{eq:SINR_final_robust} and \eqref{eq:MRC_SINR_Final}. Conventional MRC requires large arrays ($L\ge 10$) to combat low SNR and multi-antenna setups does not generate substantial improvement under noise-dominant region (e.g., $\mathrm{SNR}<-30$ dB). Meanwhile, single-antenna FAS approaches performance equvalent to $12$-antenna MRC systems with only $W=2$ and $N=100$ ports but with  less hardware overhead, equivalent to at most $L=5$ independent antenna positions.

\begin{figure}[]
\centering
\includegraphics[width=\columnwidth]{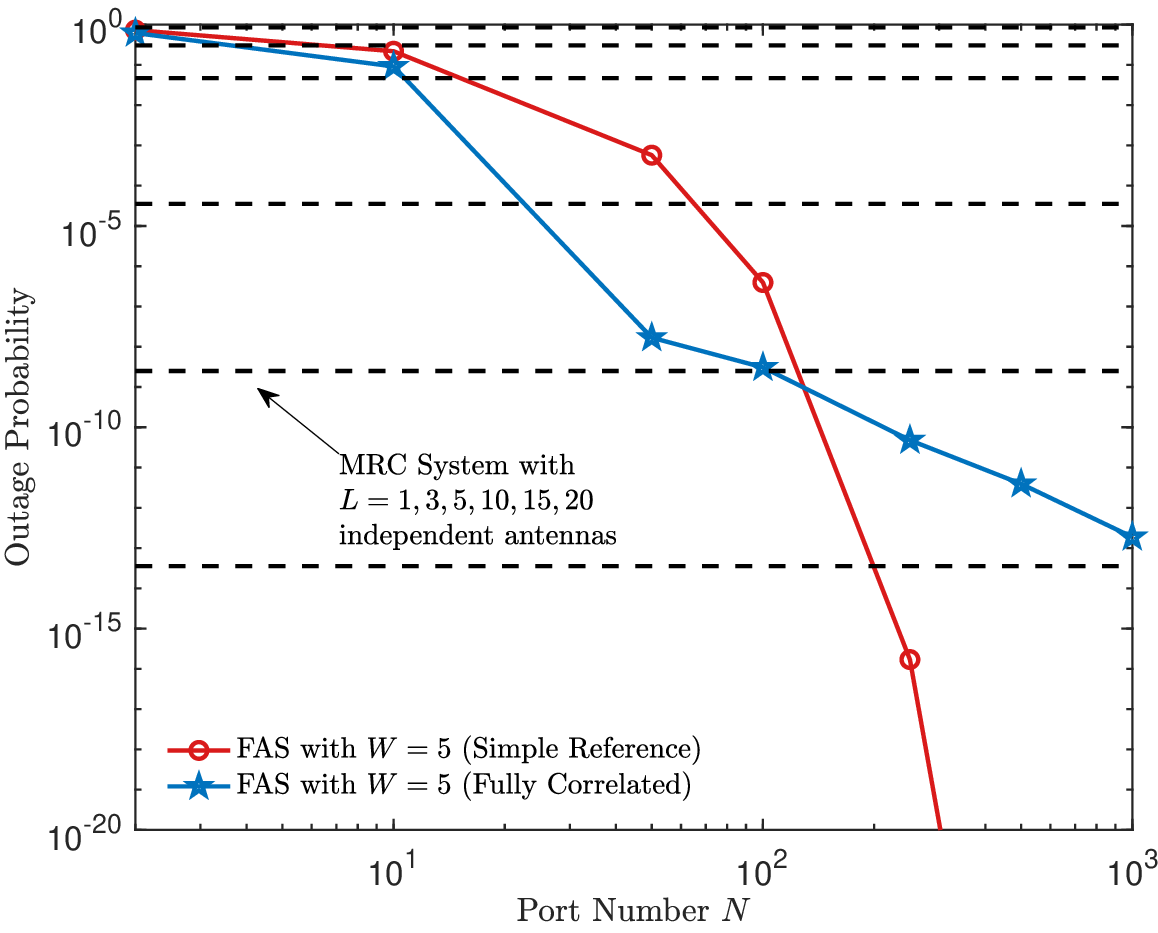}
\caption{OP against $N$ with $U=60, M=400, \mathrm{SNR}=-10$dB, $W=5, L\in\{1,3,5,10,15,20\}$.}\label{fig:OP_N}
\end{figure}

Fig.~\ref{fig:OP_N} highlights that while MRC diversity is capped by the physical size (e.g., $W=5$ supports $\approx 11$ independent antennas), FAS leverages increased $N$ to generate substantial spatial diversity within a fixed aperture. Notably, as anticipated, simplified reference model generates relatively optimistic OP performance under large $N$ region. However, both fully correlated and simplified reference models easily outperform $L=15$-MRC system with adequate number of available ports. 

\begin{figure}[]
	\centering
	\includegraphics[width=\columnwidth]{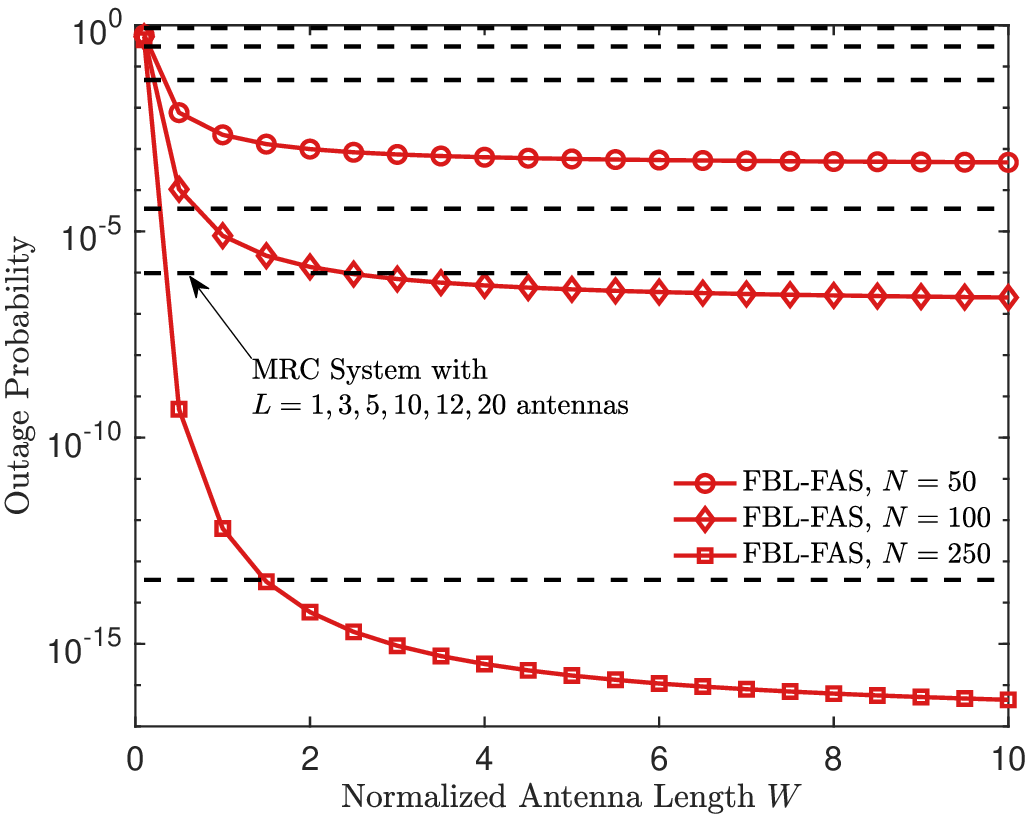}
	\caption{OP against $W$ with $U=60, M=400, \mathrm{SNR}=-10$dB, $N\in\{50,100,250\}, L\in\{1,3,5,10,12,20\}$.}\label{fig:OP_W}
\end{figure}

Fig.~\ref{fig:OP_W} further confirms that $N=100$ and $N=250$ allow FAS to approach the OP of $L=12$ and $L=20$-MRC systems, respectively. Although the OP is influenced by the tail behavior of different correlation models, the same trend is observed in all cases: $(W,N)$ determines the diversity order, and for a fixed $W$, increasing $N$ yields substantial spatial gain.

\begin{figure}[]
\centering
\includegraphics[width=\columnwidth]{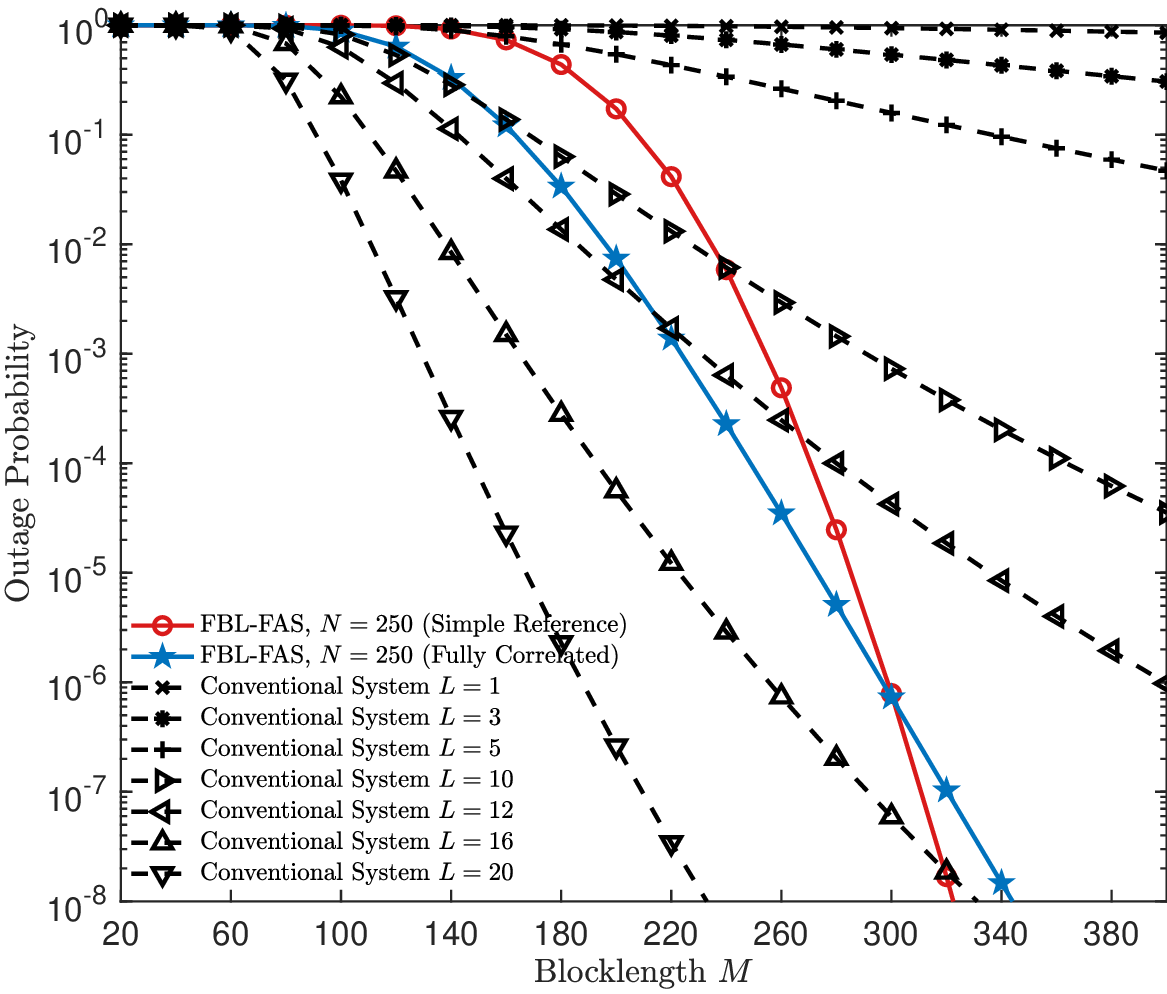}
\caption{OP against $M$ with $U=60, \mathrm{SNR}=-10$dB, $W=5$, $N=250, L\in\{1,3,5,10,12,16,20\}$.}\label{fig:OP_M}
\end{figure}

Finally, Fig.~\ref{fig:OP_M} shows a water-falling trend for the OP of FAS, dropping swiftly beyond a code rate threshold. Consistent with previous observations, conventional MRC lacks sufficient diversity at low antenna counts, whereas FAS efficiently mitigates outages even with a single RF chain regardless of what correlation modeling is considered.

\subsection{Further Observations}
A close observation of the numerical results reveals a key difference in the asymptotic behaviors of the two performance metrics. Specifically, the OP curves exhibit an error floor in the high-SNR regime ({\em because no multiple access techniques are considered and the OP becomes interference-limited}), whereas the BLER curves demonstrate a continuous waterfall characteristic without saturation. This discrepancy arises from the fundamental difference in the receiver assumptions.

Specifically, the OP analysis is predicated on the SINR expression derived in \eqref{eq:SINR_final_robust} and \eqref{eq:MRC_SINR_Final}, which essentially treats the inter-user cross-correlation as additive noise. This corresponds to a receiver employing a \textit{treat-interference-as-noise} strategy. In this formulation, increasing the transmit power scales both the desired signal power (numerator) and the interference power (denominator) proportionally. Consequently, as the thermal noise becomes negligible (i.e., $\sigma_{\eta}^2 \to 0$), the SINR saturates to a constant signal-to-interference ratio, determined solely by the codebook correlation levels. This saturation leads to the observed error floor, indicating that the system is \textit{interference-limited} under the assumption of treating interference as noise.

In contrast, the BLER upper bound derived in Theorem~\ref{Theorem 5} assumes a \textit{joint ML} detector. The joint ML detector performs decisions based on the Euclidean distance between the received vector and the superposition of all candidate codewords. Unlike linear receivers, the joint ML detector exploits the structure of the interference. As indicated by the error exponent in \eqref{eq:15}, the error probability is governed by the ratio of the Euclidean distance between codeword combinations to the thermal noise variance, i.e., $\propto \log(1 + \mathrm{SNR})$. As $\mathrm{SNR} \to \infty$, the thermal noise sphere shrinks to zero, and the detector can perfectly distinguish codeword combinations. Thus, the BLER reflects the ultimate performance limit by an optimal receiver, which remains \textit{noise-limited} rather than interference-limited.

\section{Conclusion}\label{conclusion}
This work developed an analytical framework for FASs operating in the FBL regime. A closed-form expression for BLER performance, together with a reformulated OP, is derived for a broad class of channel correlation models. The proposed BLER formulation accommodates correlation models characterized either analytically or empirically, thereby providing a unified system-level performance metric for FAS-related studies. Moreover, the outage threshold used in the OP evaluation is explicitly determined by the system configuration, enabling a more consistent assessment of reliability. Numerical results demonstrated that, even under FBL constraints, FASs can outperform conventional $L$-FPA systems employing maximal-ratio combining. Extending the present framework to fluid-antenna multiple-access systems constitutes an important direction for future research.

\begin{appendices}
\section{Proof of $\boldsymbol{c}_i^{\mathrm{H}}\boldsymbol{c}_j\sim \mathcal{CN}\left(0,M\sigma_c^4\right)$}\label{appen-a}
The $m$-th element in the $i$-th codeword distributes as circularly symmetric complex Gaussian distribution with zero mean and variance of $\sigma_c^2$ with PDF $p_{c_{i,m}}\left(x\right)=\frac{1}{\pi\sigma_c^2}e^{-\frac{|x|^2}{\sigma_c^2}}$. Thus, its conjugate follows an identical distribution, i.e., $p\left(c^{*}_{i,m}\right)=p\left(c_{i,m}\right)$. Define the product of two independent (pair selection) circularly symmetric complex Gaussian random variables as a new variable $z_m \triangleq c_{i,m}^*c_{j,m}$ with PDF \cite{Z_Distribution}
\begin{equation}
p_{z_m}(z)=\frac{2}{\pi\sigma_c^4}\,K_0\!\left(\frac{2|z|}{\sigma_c^2}\right),~z\in\mathbb{C},
\end{equation}
where $K_0(\cdot)$ is the modified Bessel function of the second kind. As $p_{z_m}(z)$ depends on $z$ only through its modulus $|z|$ and $z_m$ is circularly symmetric, $\mathrm{E}[z_m]=0$ and $\mathrm{Var}(z_m)=\mathrm{E}[|z_m|^2]$. To compute $\mathrm{E}[|z_m|^2]$, we use polar coordinates $z=\rho e^{j\phi}$, and $d^2z=\rho\,d\rho\,d\phi$ conversion to get
\begin{align}
\mathbb{E}[|z_m|^2]
&=\int_{\mathbb{C}} |z|^2 p_{z_m}(z)\,d^2z\notag\\
&=\int_{0}^{2\pi}\int_{0}^{\infty} \rho^2\left(\frac{2}{\pi\sigma_c^4}K_0\!\left(\frac{2\rho}{\sigma_c^2}\right)\right)\rho\,d\rho\,d\phi\notag\\
&=\frac{4}{\sigma_c^4}\int_{0}^{\infty}\rho^3 K_0\!\left(\frac{2\rho}{\sigma_c^2}\right)\,d\rho.\label{eq:var_step_polar}
\end{align}
Letting $u=\frac{2\rho}{\sigma_c^2}$ with derivatives $d\rho=\frac{\sigma_c^2}{2}du$, we have
\begin{align}
\int_{0}^{\infty}\rho^3 K_0\!\left(\frac{2\rho}{\sigma_c^2}\right)\,d\rho
&=\int_{0}^{\infty}\left(\frac{\sigma_c^2}{2}u\right)^3 K_0(u)\left(\frac{\sigma_c^2}{2}\,du\right)\notag\\
&=\frac{\sigma_c^8}{16}\int_{0}^{\infty}u^3K_0(u)\,du. \label{eq:var_step_sub}
\end{align}
Now, applying the standard identity $\int_{0}^{\infty} x^{\mu-1}K_{\nu}(x)\,dx=2^{\mu-2}\Gamma\!\left(\frac{\mu+\nu}{2}\right)\Gamma\!\left(\frac{\mu-\nu}{2}\right)$ with $\mu=4$ and $\nu=0$, we obtain $\int_{0}^{\infty}u^3K_0(u)\,du=2^{2}\Gamma(2)\Gamma(2)=4$, which is substituted into \eqref{eq:var_step_polar}--\eqref{eq:var_step_sub} and then yield
\begin{equation}
\mathrm{Var}(z_m)=\mathbb{E}[|z_m|^2]=\frac{4}{\sigma_c^4}\cdot\frac{\sigma_c^8}{16}\cdot 4=\sigma_c^4.
\end{equation}

Eventually, by the CLT, the sum of $M\gg 1$ random variables with identical distributions approximately follows Gaussian distribution, i.e.,  $\boldsymbol{c}_i^{\mathrm{H}}\boldsymbol{c}_j=\sum_{m=1}^{M}z_m=\sum_{m=1}^{M}c^{*}_{i,m}c_{j,m}\sim \mathcal{CN}\left(0,M\sigma_c^4\right)$, which completes the proof.

\section{Proof of $\|\boldsymbol{c}_i\|^2_2\sim~\text{Gamma distribution~} \mathcal{G}\left(M,\sigma_c^2\right)$ or $\|\boldsymbol{c}_i\|^2_2\sim \mathcal{N}\left(M\sigma_c^2,M\sigma_c^4\right)$, if $M\gg 1$}\label{appen-b}
Element-wisely, suppose the real and imaginary parts are denoted by $\mathrm{Re}\left(c_{i,m}\right)=a_{i,m},~\mathrm{Im}\left(c_{i,m}\right)=b_{i,m}$, where $a_{i,m},~b_{i,m}\sim \mathcal{N}\left(0,\frac{\sigma_c^2}{2}\right)$, $|c_{i,m}|^2=a_{i,m}^2+b_{i,m}^2$ and notably, $a_{i,m}^2,~b_{i,m}^2\sim \mathcal{G}\left(\frac{1}{2},\sigma_c^2\right)$. As a result, we have
\begin{equation}
|c_{i,m}|^2\sim \mathcal{G}\left(\frac{1}{2}+\frac{1}{2},\sigma_c^2\right)=\mathcal{E}\left(\frac{1}{\sigma_c^2}\right),
\end{equation}
where $\mathcal{E}\left(\cdot\right)$ denote exponential distribution with PDF
\begin{equation}
p_{|c_{i,m}|^2}\left(x\right) = \frac{1}{\sigma_c^2}e^{-\frac{x}{\sigma_c^2}},~x\ge 0,
\end{equation}
whose statistics are $\mathrm{E}\left[|c_{i,m}|^2\right]=\sigma_c^2$ and $\mathrm{Var}\left[|c_{i,m}|^2\right]=\sigma_c^4$.
\par\indent For inner product/$l_2$-norm, $\|\boldsymbol{c}_i\|_2^2=\sum_{m=1}^{M}|c_{i,m}|^2$, where $|c_{i,m}|^2,~m=1,\ldots,M$ are independent. Hence, $\|\boldsymbol{c}_i\|_2^2$ distributes as $\mathcal{G}\left(M,\sigma_c^2\right)$ with statistics $\mathrm{E}\left[\|\boldsymbol{c}_i\|_2^2\right]=M\sigma_c^2$ and $\mathrm{Var}\left[\|\boldsymbol{c}_i\|_2^2\right]=M\sigma_c^4$, where $X\sim\mathcal{G}(k,\theta)$ means $f_X(x)=\frac{1}{\Gamma(k)\theta^k}x^{k-1}e^{-x/\theta},x\geq0$. Furthermore, by the CLT, if $M\gg 1$, $\frac{1}{M}\sum_{m=1}^{M}\lvert c_{i,m}\rvert^2 \sim \mathcal{N}\!\left(\mathrm{E}\!\left\{\lvert c_{i,m}\rvert^2\right\},\operatorname{Var}\!\left(\lvert c_{i,m}\rvert^2\right)/M\right)$, which is equivalent to $\lVert c_i\rVert_2^2=\sum_{m=1}^{M}\lvert c_{i,m}\rvert^2\sim\mathcal{N}\!\left(M\mathbb{E}\!\left\{\lvert c_{i,m}\rvert^2\right\},M\operatorname{Var}\!\left(\lvert c_{i,m}\rvert^2\right)\right).$ After substituting the expectation $\mathrm{E}\!\left[\lvert c_{i,m}\rvert^2\right]=\sigma_c^2$ and the variance $\operatorname{Var}\!\left[\lvert c_{i,m}\rvert^2\right]=\sigma_c^4,$ we could have $ \lVert c_i\rVert_2^2 \approx\mathcal{N}\!\left(M\sigma_c^2,M\sigma_c^4\right),$ which completes the proof of Theorem~\ref{Theorem 2}.

\section{Derivation of $\rho_{\max}$ in \eqref{eq:10}}\label{appen-c}
To prove the result, we first explain the amplitude distribution PDF of the inner product of two {\em normalized} codewords, which is Rayleigh distributed, indicating the PDF tail behaviors of $T=\frac{U\left(U-1\right)}{2}$ potential $\rho_{i,j}$ in \eqref{eq:7b} are exponentially decaying. Consequently, the distribution of $\rho_{\max}  = \max_{i\neq j}\rho_{i,j}$ can be well described by Gumbel distribution. Thus, we derive the analytical estimation under FBL for expectation of maximum correlation $\bar{\rho}_{\max}=\mathrm{E}\left[\rho_{\max}\right]=\mathrm{E}\left[\max_{i\neq j}\rho_{i,j}\right]$. Here, we list some remarks useful in the sequel.

\begin{remark}\label{Remark C1}
\textit{For Gumbel distribution \cite{Gumel-1-Extreme_value, Gumbel-2,Gumbel-3} denoted by $\mathscr{G}_Z\left(a,b\right)$ where $a$ is the location parameter and $b>0$ is the scale parameter, the general PDF is given by
\begin{equation}
p_Z(a,b) = \frac{1}{b}e^{\left(-\frac{z-a}{b}-e^{-\frac{z-a}{b}}\right)},
\end{equation}
and the CDF is
\begin{equation}
C_Z\left(a,b\right)=e^{\left(-e^{-\frac{z-a}{b}}\right)}.
\end{equation}
Also, $\mathrm{E}\left[Z\right]=a+b\gamma$ and $\mathrm{Var}\left[Z\right]=\frac{\pi^2b^2}{6}$ where $\gamma\approx 0.57721$ denotes the Euler-Mascheroni constant.}
\end{remark}

\begin{remark}\label{Remark C2}
\textit{If $a=0,b=1$, one will have a standard Gumbel distribution $\mathscr{G}_Z\left(0,1\right)$ whose CDF is $C_Z(0,1) = e^{\left(-e^{-z}\right)}$ with expectation $\mathrm{E}\left[Z\right]=\gamma$. Moreover, the value of standard Gumbel CDF at central point, i.e., $z=0$, is a fixed constant.}
\end{remark}

Defining the normalized codewords $\boldsymbol{n}_i=\frac{\boldsymbol{c}_i}{\|\boldsymbol{c}_i\|_2}$, based on Corollary~\ref{Corollary 2} in Section~\ref{Main Results-A}, it is known that $\boldsymbol{n}_i^{\mathrm{H}}\boldsymbol{n}_j\sim \mathcal{CN}\left(0,\frac{1}{M}\right)$. Therefore, the amplitude/modulus of $\boldsymbol{n}_i^{\mathrm{H}}\boldsymbol{n}_j$, i.e., $\rho_{i,j}=|\boldsymbol{n}_i^{\mathrm{H}}\boldsymbol{n}_j|$, abides by Rayleigh distribution with PDF
\begin{equation}\label{pdf_rho}
p_{\rho_{i,j}}\left(x\right)=2Mx\cdot e^{-Mx^2},~x\ge0,
\end{equation}
and has the corresponding CDF given by
\begin{equation}\label{cdf_rho}
C_{\rho_{i,j}}\left(x\right)=1-e^{-Mx^2},~x\ge0.
\end{equation}
Note that the tail behaviors of \eqref{pdf_rho} and \eqref{cdf_rho} are {\em exponentially decaying}, which explains why Gumbel distribution is appropriate to describe the extreme value distribution in \eqref{eq:7b} \cite{Gumel-1-Extreme_value}.

Now, considering $T=\frac{U\left(U-1\right)}{2}$ independent potential correlation coefficients and their CDF in \eqref{cdf_rho}, the joint probability of event $\rho_{\max} = \max_{i\neq j}\rho_{i,j}$ equals to
\begin{equation}\label{joint_probability}
P\left(\rho_{\max}\le x\right) = \left(1-e^{-Mx^2}\right)^{T}.
\end{equation}
Since the tail behavior of Rayleigh distribution is exponentially decaying, Gumbel distribution is a suitable tool to derive analytical results for extreme value of correlation coefficients. 
In the following, we derive the asymptotic distribution of $\rho_{\max}$ and its expectation $\bar{\rho}_{\max}=\mathrm{E}\left[\rho_{\max}\right]$ in three steps:
\begin{itemize}
	\item[--] {\em Variable Transformation:} To cast \eqref{joint_probability} into the limit structure $\left(1-\frac{c}{T}\right)^{T}\to e^{-c}$, the exponential term should satisfy $e^{-Mx^2}=\frac{e^{-z}}{T}$, i.e., $Mx^2=\ln T+z$. This motivates defining the random variable
	\begin{equation}\label{Z_def}
		Z \triangleq M\rho_{\max}^2-\ln T,
	\end{equation}
	which is a monotone one-to-one transformation of $\rho_{\max}$ for $\rho_{\max}\ge 0$. Then, the CDF of $Z$ follows \emph{exactly} from \eqref{joint_probability} as
	\begin{align}\label{exchanged_eq}
		P\left(Z\le z\right) &= P\left(\rho_{\max}\le \sqrt{\tfrac{\ln T+z}{M}}\right)
		\\ \nonumber &= \left(1-e^{-\left(\ln T+z\right)}\right)^{T}
		\\ \nonumber &= \left(1-\frac{e^{-z}}{T}\right)^{T}.
	\end{align}
	Since $T=\frac{U(U-1)}{2}$ grows quadratically and easily satisfies $T\gg 1$, applying $\left(1-\frac{c}{T}\right)^{T}\approx e^{-c}$ yields
	\begin{equation}\label{Gumbel-converted}
		P\left(Z\le z\right)\approx e^{-e^{-z}},
	\end{equation}
	i.e., $Z$ asymptotically follows the standard Gumbel distribution $\mathscr{G}_z(0,1)$ in Remark~\ref{Remark C2}. Note that no approximation other than the large-$T$ limit is involved so far.
	
	\item[--] {\em Asymptotic Linearization:} Inverting \eqref{Z_def} gives the exact relation
	\begin{equation}\label{inverse_map}
		\rho_{\max}=\sqrt{\frac{\ln T+Z}{M}}=\sqrt{\frac{\ln T}{M}}\sqrt{1+\frac{Z}{\ln T}}.
	\end{equation}
	By Remark~\ref{Remark C1} and Remark~\ref{Remark C2}, $Z$ has bounded mean $\gamma$ and variance $\frac{\pi^2}{6}$, both independent of $T$, whereas $\ln T\gg 1$ or even  $\ln T\rightarrow \infty$ under multiple/massive uplink transmission. Hence $\frac{Z}{\ln T}\ll 1$ holds with high probability, and the first-order Taylor expansion $\sqrt{1+u}\approx 1+\frac{u}{2}$ applied to \eqref{inverse_map} gives
	\begin{equation}\label{linearized}
		\rho_{\max}\approx \underbrace{\sqrt{\frac{\ln T}{M}}}_{\triangleq\, b_z\ (\text{location})}+\underbrace{\frac{1}{2\sqrt{M\ln T}}}_{\triangleq\, a_z\ (\text{scale})}Z.
	\end{equation}
	That is, $\rho_{\max}$ is (asymptotically) an affine transformation of a standard Gumbel variable, with location parameter $b_z$ and scale parameter $a_z$ emerging directly from the expansion. This is consistent with the classical extreme value theory, where the normalizing constants of the maximum of Rayleigh-tailed variables are obtained via such linearization \cite{Gumel-1-Extreme_value,Gumbel-3}.
	
	\item[--] {\em Maximum Correlation Expectation:} Taking expectation on both sides of \eqref{linearized} and using $\mathrm{E}[Z]=\gamma$ from Remark~\ref{Remark C2}, we obtain
	\begin{equation}
		\begin{aligned}
					\bar{\rho}_{\max}&=\mathrm{E}\left[\rho_{\max}\right] \\
					&\approx b_z+\gamma a_z= \sqrt{\frac{\ln T}{M}}+\gamma\frac{1}{2\sqrt{M\ln T}},
		\end{aligned}
	\end{equation}
	which completes the proof for Theorem~\ref{Theorem 4} in \eqref{eq:10}.
\end{itemize}

\section{BLER Conditioned on $|g_{\mathrm{FAS}}|$}\label{appen-d}
For the $m$-th element $\eta_m'$ in $\boldsymbol{\eta}'=\left(\boldsymbol{X}_{{U}'}-{\boldsymbol{X}}'\right){\boldsymbol{g}}'$, $\eta_m'=\sum_{u=1}^{{U}'}\left(c_{u_i,m}-c_{u_j,m}\right)g_{\mathrm{FAS}}$ where $c_{u_i,m}\neq c_{u_j,m}$ are elements from any two different codewords. With $c_{u_i,m},c_{u_j,m}\sim \mathcal{CN}\left(0,\sigma_c^2\right)$ being independent, the distribution conditioned on $|g_{\mathrm{FAS}}|$ is $\eta_m'\sim \mathcal{CN}\Bigg(0,\underbrace{2{U}'\sigma_c^2|g_{\mathrm{FAS}}|^2}_{\sigma^2_{\eta'}} \Bigg\vert |g_{\mathrm{FAS}}|\Bigg)$.

Now, conditioned on $|g_{\mathrm{FAS}}|$, we have
\begin{align}\label{BLER-conditioned}
P\left(\mathcal{E}_{{\mathcal{U}}'}\Big\vert |g_{\mathrm{FAS}}|\right) =P\left(\|\boldsymbol{\eta}'+\boldsymbol{\eta}\|^2_2 < \|\boldsymbol{\eta}\|^2_2 \Big\vert |g_{\mathrm{FAS}}| \right).
\end{align}
For conciseness, the condition notation
$\left(\left(\cdot\right)\Big\vert |g_{\mathrm{FAS}}|\right)$ is omitted
unless particularly stated. Define $X\triangleq\|\boldsymbol{\eta}\|_2^2-\|\boldsymbol{\eta}'+\boldsymbol{\eta}\|_2^2.$
By utilizing the Chernoff inequality/bound $P(X>0)\leq\mathrm{E}\!\left[e^{\lambda_1 X}\right],~ \lambda_1>0,$
\cite{chernoff3,chernoff2,chernoff1}, \eqref{BLER-conditioned} can be upper-bounded as
\begin{subequations}\label{chernoff}
	\begin{align}
		&P\left(
		\|\boldsymbol{\eta}'+\boldsymbol{\eta}\|_2^2
		<
		\|\boldsymbol{\eta}\|_2^2
		\right)
		\leq
		\mathrm{E}\left[
		e^{-\lambda_1\|\boldsymbol{\eta}'+\boldsymbol{\eta}\|_2^2
			+\lambda_1\|\boldsymbol{\eta}\|_2^2}
		\right]
		\notag\\
		&=
		\mathrm{E}\left[
		\frac{
			e^{\lambda_1\|\boldsymbol{\eta}\|_2^2}
			e^{\frac{-\lambda_1\|\boldsymbol{\eta}\|_2^2}
				{1+\lambda_1\sigma_{\eta'}^2}}
		}{
			\left(1+\lambda_1\sigma_{\eta'}^2\right)^M
		}
		\right]
		=
		\mathrm{E}\left[
		\frac{
			e^{\left(
				\frac{\lambda_1^2\sigma_{\eta'}^2}
				{1+\lambda_1\sigma_{\eta'}^2}
				\right)\|\boldsymbol{\eta}\|_2^2}
		}{
			\left(1+\lambda_1\sigma_{\eta'}^2\right)^M
		}
		\right]
		\label{chernoff-b}\\
		&=
		\frac{1}{\left(1+\lambda_1\sigma_{\eta'}^2\right)^M}
		\frac{1}{
			\left(
			1-
			\frac{
				\lambda_1^2\sigma_{\eta'}^2\sigma_{\eta}^2/M
			}{
				1+\lambda_1\sigma_{\eta'}^2
			}
			\right)^M
		}
		\label{chernoff-c}\\
		&=
		\frac{1}{
			\left(
			1+\lambda_1\sigma_{\eta'}^2
			-\lambda_1^2\sigma_{\eta'}^2\sigma_{\eta}^2/M
			\right)^M
		},
		\label{chernoff-d}
	\end{align}
	where the expectation is finite for any $\lambda_1>0$ satisfying $1+\lambda_1\sigma_{\eta'}^2-\frac{\lambda_1^2\sigma_{\eta'}^2\sigma_{\eta}^2}{M}>0.$ In the derivation from \eqref{chernoff-b} to \eqref{chernoff-c}, the
	identity $\mathrm{E}\left[e^{x\|\boldsymbol{a}+\boldsymbol{b}\|_2^2}\right]=\frac{ e^{\frac{x\|\boldsymbol{b}\|_2^2}{1-x\sigma_a^2}} }{ \left(1-x\sigma_a^2\right)^M},$ in which the entries of $\boldsymbol{a}$ follow $\mathcal{CN}(\mathbf{0},\sigma_a^2)$, has been adopted twice.
	Since the Chernoff inequality holds for every admissible
	$\lambda_1>0$, the tightest upper bound obtainable from
	\eqref{chernoff-d} is found by minimizing its right-hand side over
	$\lambda_1$. To this end, define its denominator base as $
	q(\lambda_1) \triangleq 1+\lambda_1\sigma_{\eta'}^2-\frac{\lambda_1^2\sigma_{\eta'}^2\sigma_{\eta}^2}{M}.$ Because the right-hand side of \eqref{chernoff-d} is $q(\lambda_1)^{-M}$, minimizing the upper bound is equivalent to maximizing $q(\lambda_1)$. Moreover, $q'(\lambda_1)=\sigma_{\eta'}^2-\frac{2\lambda_1\sigma_{\eta'}^2\sigma_{\eta}^2}{M},~q''(\lambda_1)=-\frac{2\sigma_{\eta'}^2\sigma_{\eta}^2}{M}<0.$ Thus, $q(\lambda_1)$ is strictly concave and its unique maximum is achieved at $\lambda_1^\star=\frac{M}{2\sigma_{\eta}^2}=\frac{0.5M}{\sigma_{\eta}^2}.$ Substituting $\lambda_1^\star$ into \eqref{chernoff-d} yields
	\begin{align}
		&P\left(
		\|\boldsymbol{\eta}'+\boldsymbol{\eta}\|_2^2
		<
		\|\boldsymbol{\eta}\|_2^2
		\right)
		\leq
		e^{-M\log\left(
			1+\frac{0.25M\sigma_{\eta'}^2}{\sigma_{\eta}^2}
			\right)}.
		\label{chernoff-e}
	\end{align}
\end{subequations}

Consequently, considering the error summation in \eqref{eq:14}, the BLER conditioned on
$|g_{\mathrm{FAS}}|$ is upper-bounded by
\begin{equation}
P\left( \mathcal{E}_{{\mathcal{U}}'} \Big\vert |g_{\mathrm{FAS}}| \right)\leq \sum_{{U}'=0}^{U}\frac{{U}'}{U}e^{{L}'-M\log\left(1+\frac{0.25M\sigma_{\eta'}^2}{\sigma_{\eta}^2}\right)},
\end{equation}
where ${L}'=\log\binom{U}{{U}'}^2=\sum_{i=1}^{{U}'-1}2\log\frac{U-i}{{U}'-i},~\sigma_{\eta'}^2=2{U}'\sigma_c^2|g_{\mathrm{FAS}}|^2,$ and the ratio $\frac{{U}'}{U}$ denotes the BLER when ${U}'$ errors exist, which completes the proof of Theorem~\ref{Theorem 5}.

\begin{figure}[]
\centering
\includegraphics[width=\columnwidth]{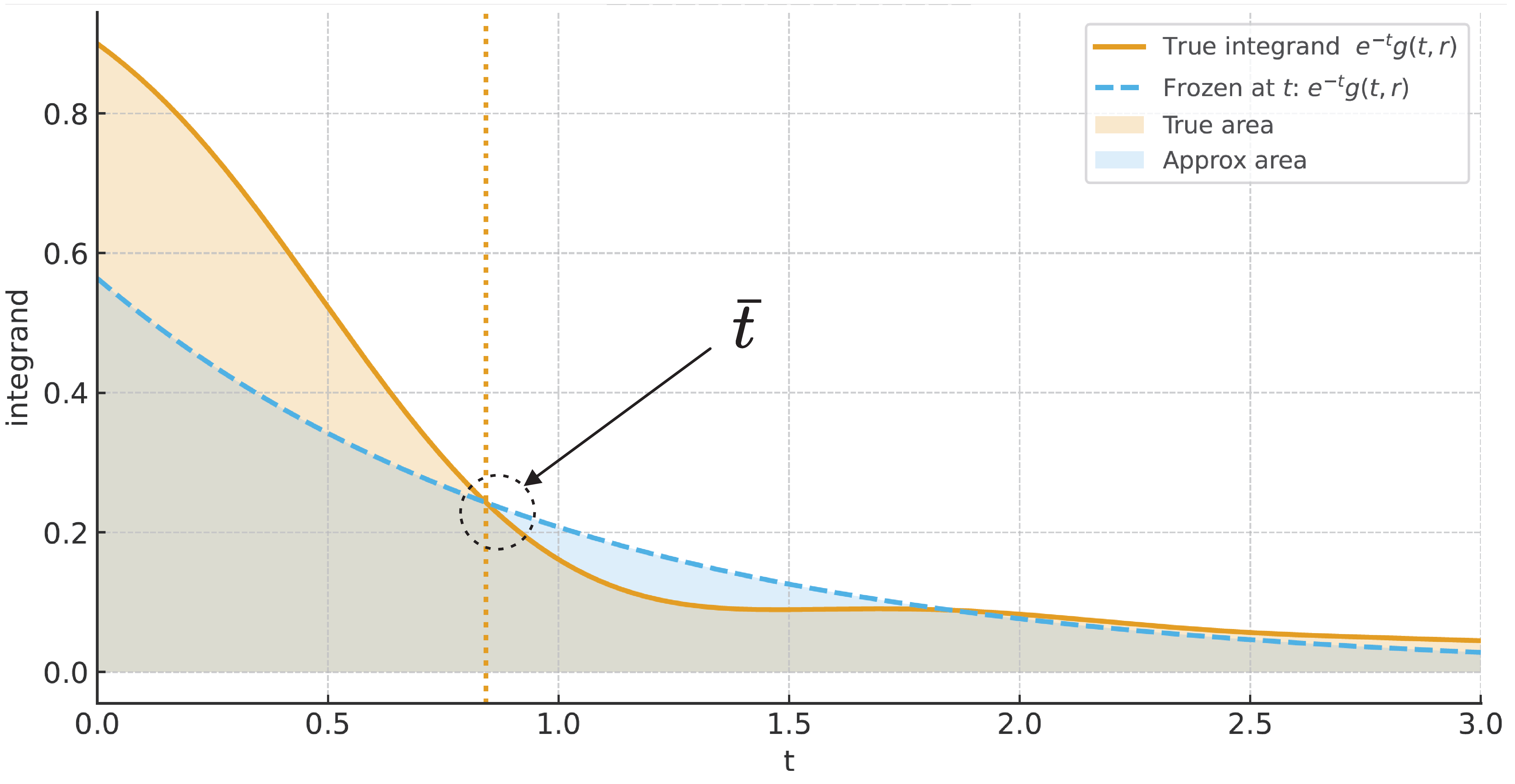}
\caption{Illustration of Lemma~\ref{Lemma 1} in \eqref{eq:lemma 1} where point $\bar{t}$ resembles the area of the two shaded planes.}\label{fig:frozen_point_approx}
\end{figure}

\section{Derivations of the PDF $f_{|g_{\mathrm{FAS}}|}\left(r\right)$}\label{appen-e}
With the conditional BLER in Appendix \ref{appen-d}, we can finalize the upper-bounding on BLER for FBL-FAS with the PDF of $|g_{\mathrm{FAS}}|$. To proceed, we find the following remarks useful.

\begin{remark}\label{Remark E1}
\textit{In \cite{fas-twc-21}, the joint PDF of $|g_1|,|g_2|,\ldots,|g_N|$ is}
\begin{equation}
\begin{aligned}
&p_{|g_1|,\ldots,|g_N|}\left(r_1,\ldots,r_N\right)\\
&=\prod_{k=1}^N \frac{2 r_k}{\sigma^2 (1 - \mu_k^2)} e^{ \left( -\frac{r_k^2 + \mu_k^2 r_1^2}{\sigma^2 (1 - \mu_k^2)} \right)} I_0\left( \frac{2 \mu_k r_1 r_k}{\sigma^2 (1 - \mu_k^2)} \right),
\end{aligned}
\end{equation}
\textit{and the joint CDF of $|g_1|,|g_2|,\ldots,|g_N|$ is given by}
\begin{equation}\label{eq:joint_cdf_appedx}
\begin{aligned}
&C_{|g_1|, \dots, |g_N|}(r_1, \dots, r_N)\\
&= \int_0^{\frac{r_1^2}{\sigma^2}} e^{-t} \times\\
& \quad \prod_{k=2}^N \left[ 1 - Q_1\left( \sqrt{\frac{2 \mu_k^2}{1 - \mu_k^2}} \sqrt{t}, \sqrt{\frac{2}{\sigma^2 (1 - \mu_k^2)}} r_k \right) \right] dt,
\end{aligned}
\end{equation}
\textit{where $I_0\left(\cdot\right)$ is the zero-order modified Bessel function of the first kind and $Q_1\left(a,b\right)$ is the first-order Marcum $Q$-function.}
\end{remark}

With Remark~\ref{Remark E1}, the CDF of $|g_{\mathrm{FAS}}|$ becomes
\begin{equation}\label{joint CDF}
\begin{aligned}
&C_{|g_{\mathrm{FAS}}|}(r) = P(|g_1| \leq r, \dots, |g_N| \leq r)\\
& = C_{|g_1|, \dots, |g_N|}(r_1=r, \dots, r_N=r)\\
& = \int_0^{\frac{r^2}{\sigma^2}} e^{-t} \prod_{k=2}^N \left[ 1 - Q_1\left( \sqrt{\frac{2 \mu_k^2}{1 - \mu_k^2}} \sqrt{t}, \sqrt{\frac{2}{\sigma^2 (1 - \mu_k^2)}} r \right) \right] \mathrm{d}t.
\end{aligned}
\end{equation}
As a result, the $|g_{\mathrm{FAS}}|$'s PDF is given by
\begin{equation}
f_{|g_{\mathrm{FAS}}|}\left(r\right)=\frac{\mathrm{d}}{\mathrm{d}r} C_{|g_{\mathrm{FAS}}|}(r).
\end{equation}
Utilizing Leibniz rule of $\frac{\mathrm{d}}{\mathrm{d}r} \int_{a(r)}^{b(r)} g(t, r) \, \mathrm{d}t = g(b(r), r) \frac{\mathrm{d}b}{\mathrm{d}r} - g(a(r), r) \frac{\mathrm{d}a}{\mathrm{d}r} + \int_{a(r)}^{b(r)} \frac{\partial g(t, r)}{\partial r} \, \mathrm{d}t$ with $a(r)=0,b(r)=\frac{r^2}{\sigma^2}$, and substituting $u = \frac{r^2}{\sigma^2},\frac{du}{dr} = \frac{2r}{\sigma^2}$, one gets the PDF in \eqref{d_FAS_PDF} (see top of next page).
\begin{figure*}[]
\begin{equation}\label{d_FAS_PDF}
\begin{aligned}
&f_{|g_{\mathrm{FAS}}|}(r)=\\
 &\frac{2r}{\sigma^2} e^{-\frac{r^2}{\sigma^2}} \prod_{k=2}^N \left[ 1 - Q_1\left( \sqrt{\frac{2 \mu_k^2}{1 - \mu_k^2}} \frac{r}{\sigma}, \sqrt{\frac{2}{1 - \mu_k^2}} \frac{r}{\sigma} \right) \right] +\sum_{i=2}^N \int_0^{\frac{r^2}{\sigma^2}} e^{-t} \left( \prod_{k=2, k \neq i}^N \left[ 1 - Q_1\left( \sqrt{\frac{2 \mu_k^2}{1 - \mu_k^2}} \sqrt{t}, \sqrt{\frac{2}{1 - \mu_k^2}} \frac{r}{\sigma} \right) \right] \right) \times\\
& \underbrace{\frac{\partial}{\partial r} \left[ 1 - Q_1\left( \sqrt{\frac{2 \mu_i^2}{1 - \mu_i^2}} \sqrt{t}, \sqrt{\frac{2}{1 - \mu_i^2}} \frac{r}{\sigma} \right) \right] \mathrm{d}t}_{\eqref{d_component}},
\end{aligned}
\end{equation}
\hrulefill
\end{figure*}

\begin{figure*}[]
\begin{equation}\label{d_component}
\begin{aligned}
\frac{\partial}{\partial r} \left[ 1 - Q_1\left( \sqrt{\frac{2 \mu_i^2}{1 - \mu_i^2}} \sqrt{t}, \sqrt{\frac{2}{1 - \mu_i^2}} \frac{r}{\sigma} \right) \right]= \frac{2r}{1 - \mu_i^2} \frac{1}{\sigma^2} \left(e^{ -\frac{\frac{2 \mu_i^2 t}{1 - \mu_i^2} + \frac{2 r^2}{\sigma^2 (1 - \mu_i^2)}}{2} }\right)\times I_0\left( \sqrt{\frac{2 \mu_i^2 t}{1 - \mu_i^2}} \cdot \sqrt{\frac{2}{1 - \mu_i^2}} \frac{r}{\sigma} \right),
\end{aligned}
\end{equation}
\hrulefill
\end{figure*}

Consider the entities (proof in Appendix~\ref{ID_Proof}): $\frac{\partial}{\partial d} Q_1(c, d) = -d e^{\left( -\frac{c^2 + d^2}{2} \right)} I_0(cd)$, $\frac{\partial}{\partial r} Q_1(a \sqrt{t}, b r) = -b^2 e^{\left( -\frac{a^2 t + b^2 r^2}{2} \right)} I_0(a \sqrt{t} b r)$ and substituting $ a = \sqrt{\frac{2 \mu_i^2}{1 - \mu_i^2}}$, $ b = \sqrt{\frac{2}{1 - \mu_i^2}} \frac{1}{\sigma}$, \eqref{d_component} is subsequently obtained.
Finally, substituting \eqref{d_component} into \eqref{d_FAS_PDF} concludes the PDF in \eqref{eq:16}.

\section{MVTI-based CDF/PDF Simplification}\label{appen-f}
\begin{lemma}\label{Lemma 1}
\textit{The integral $F=\int_{0}^{a_r}e^{-t}g\left(t,r\right)\mathrm{d}t$, if $g\left(t,r\right)$ is continuous, and it can be approximated as}  
\begin{equation}\label{eq:lemma 1}
F\approx g\left(\bar{t},r\right)\int_{0}^{a_r}e^{-t}\mathrm{d}t=\left(1-a_r\right)g\left(\bar{t},r\right),
\end{equation}
\textit{where $a_r$ may be related to $r$ and $\bar{t}=\frac{1-\left(1+a_r\right)e^{-a_r}}{1-e^{-a_r}}$ is termed as the `frozen' point ensuring the equity.}
\end{lemma}

\begin{IEEEproof}
See Appendix~\ref{appen-g}.
\end{IEEEproof}

To explain the meaning of the frozen point $\bar{t}$, an example is provided in Fig.~\ref{fig:frozen_point_approx}, where $\bar{t}$ represents the point resembling the area of two shaded regions and approximates the integral value of two functions. By applying Lemma~\ref{Lemma 1} in \eqref{eq:lemma 1}, the CDF in \eqref{joint CDF} and the PDF in \eqref{d_FAS_PDF} can be simplified into the compact form in \eqref{simplifed_pdf_cdf}, avoiding the  product-integral calculation required previously. This completes the proof of Theorem~\ref{Theorem 7}.

\section{Proof on Lemma~\ref{Lemma 1}}\label{appen-g}
Based on the MVTI, there exists a point $\bar{t}$ such that  
\begin{equation}
F=\int_{0}^{a_r} e^{-t} g(t,r)\,\mathrm{d}t = g(\bar{t},r)\int_{0}^{a_r} e^{-t}\,\mathrm{d}t.  
\end{equation}
To estimate $\bar{t}$, we apply a Taylor expansion of $g(t,r)$. Considering the expansion around $t=0$, we have  
\begin{align}
g(t,r) &\approx g(0,r) + g^{\prime}(0,r)t + \tfrac{1}{2} g^{\prime\prime}(0,r)t^2 + \cdots, \\
g(\bar{t},r) &\approx g(0,r) + g^{\prime}(0,r)\bar{t} + \tfrac{1}{2} g^{\prime\prime}(0,r)\bar{t}^2 + \cdots.
\end{align}
These expansions are then substituted into the left- and right-hand sides of the integral, respectively:
\begin{equation}
\begin{aligned}
F&=\underbrace{g\left(0,r\right)\int_{0}^{a_r}e^{-t}\mathrm{d}t+g^{\prime}\left(0,r\right)\int_{0}^{a_r}te^{-t}\mathrm{d}t+\ldots}_{=g\left(0,r\right)\left(1-e^{-a_r}\right)+g^{\prime}\left(0,r\right)\left(1-\left(a_r+1\right)e^{-a_r}\right)+\ldots}\\
&=(1-e^{-a_r})\left[g(0,r)+g^{\prime}(0,r)\bar{t}+\frac{1}{2}g^{\prime\prime}(0,r)\bar{t}^2+\ldots\right],
\end{aligned}
\end{equation}
where the first-order term should be identical. Thus, we have
\begin{equation}
\begin{aligned}
g^{\prime}\left(0,r\right)\left(1-\left(a_r+1\right)e^{-a_r}\right)&=(1-e^{-a_r})g^{\prime}(0,r)\bar{t}\\
\rightarrow \bar{t}&=\frac{1-\left(1+a_r\right)e^{-a_r}}{1-e^{-a_r}},
\end{aligned}
\end{equation}
which completes the proof.

\section{Derivation of OP for FBL-FAS}\label{appen-h}
Given the signal model in \eqref{eq:5} where $\boldsymbol{y}= g_{i,k}\boldsymbol{x}_i+\sum_{u\neq i}^{U}g_{u,k}\boldsymbol{x}_u+\boldsymbol{\eta}$. We apply a linear equalizer $g_{\mathrm{eq}}$ to obtain the estimated signal $\tilde{\boldsymbol{x}}_i = g_{\mathrm{eq}}\boldsymbol{y}=g_{\mathrm{eq}}g_{i,k}\boldsymbol{x}_i+\sum_{u\neq i}^{U}g_{\mathrm{eq}}g_{u,k}\boldsymbol{x}_u+g_{\mathrm{eq}}\boldsymbol{\eta}$. The average SINR, $\Gamma$, is defined as
\begin{equation}\label{eq:SINR_def_robust}
\Gamma=\frac{P_{S}}{P_{I}+P_{N}},
\end{equation}
where $P_S$ denotes the average signal power, $P_I$ is the interference power and $P_N$ is the average noise power. Note that the average interference power $P_{I}$ requires the non-zero codebook correlations derived in Theorem~\ref{Theorem 3} and Theorem~\ref{Theorem 4}. We outline the derivation steps below for obtaining these power values.
\begin{itemize}
\item $P_S$: Conditioned on the channel realization $g_{i,k}$ and assuming unit-energy codewords, i.e., $\|\boldsymbol{x}_i\|_2^2=1$, the power of the desired signal is given by
\begin{equation}\label{eq:Ps_robust}
P_{S} = |g_{\mathrm{eq}}|^2 |g_{i,k}|^2 \|\boldsymbol{x}_i\|_2^2 = |g_{\mathrm{eq}}|^2 |g_{i,k}|^2.
\end{equation}
\item $P_I$: The interference power is the squared magnitude of the summed interfering signals. Since orthogonal codeword under FBL is usually not valid, we retain the cross-terms due to the finite codebook correlation, giving
\begin{equation}\label{eq:Pi_expansion}
\begin{aligned}
&P_{I} = \left\| g_{\mathrm{eq}} \sum_{u\neq i}^{U} g_{u,k} \boldsymbol{x}_u \right\|^2 \\
&= |g_{\mathrm{eq}}|^2 \left( \sum_{u\neq i}^{U} |g_{u,k}|^2 \|\boldsymbol{x}_u\|_2^2 + \sum_{u\neq i}^{U} \sum_{j\neq i\atop j\neq u}^{U} g_{u,k} g_{j,k}^* \boldsymbol{x}_j^{\mathrm{H}} \boldsymbol{x}_u \right).
\end{aligned}
\end{equation}
Using the definition of the correlation coefficient $\rho_{j,u} = |\boldsymbol{x}_j^{\mathrm{H}} \boldsymbol{x}_u|$ (for unit-norm codewords) and using the approximation based on the averaged correlation $\bar{\rho}$ in Theorem~\ref{Theorem 3} to estimate the magnitude of the cross-terms:
\begin{equation}\label{eq:Pi_approx}
\begin{aligned}
P_{I} &\le |g_{\mathrm{eq}}|^2 \left( \sum_{u\neq i}^{U} |g_{u,k}|^2 + \sum_{u\neq i}^{U} \sum_{j\neq i\atop j\neq u}^{U} |g_{u,k}| |g_{j,k}| \rho_{j,u} \right) \\
&\approx |g_{\mathrm{eq}}|^2 \left( \sum_{u\neq i}^{U} |g_{u,k}|^2 + \bar{\rho} \sum_{u\neq i}^{U} \sum_{j\neq i\atop j\neq u}^{U} |g_{u,k}| |g_{j,k}| \right)\\
&\approx (U-1)|g_{\mathrm{eq}}|^2\sigma^2 + (U-1)(U-2) |g_{\mathrm{eq}}|^2\frac{\pi}{4}\sigma^2 \bar{\rho},
\end{aligned}
\end{equation}
in which we have used the inequality $\sum a_i\leq\sum|b_i|$, Theorem~\ref{Theorem 3} and the approximation consideing that $ |g_{u,k}|$, $|g_{j,k}|$ are replaced by their statistical expectations with both following Rayleigh distribution with variance $\sigma^2$.
\item $P_N$: The post-equalization noise power is given by
\begin{equation}\label{eq:Pn_robust}
P_{N} = \mathrm{E}\left[\|g_{\mathrm{eq}}\boldsymbol{\eta}\|_2^2\right]=\mathrm{tr}\left(|g_{\mathrm{eq}}|^2\frac{\sigma^2_{\eta}}{M}\boldsymbol{I}_{M}\right)=|g_{\mathrm{eq}}|^2 \sigma^2_{\eta}.
\end{equation}
\end{itemize}

Substituting \eqref{eq:Ps_robust}, \eqref{eq:Pi_approx}, and \eqref{eq:Pn_robust} into \eqref{eq:SINR_def_robust}, the scalar equalizer gain $|g_{\mathrm{eq}}|^2$ cancels out. The resulting instantaneous SINR expression, accounting for codebook correlation, is
\begin{equation}\label{eq:SINR_final_robust}
\begin{aligned}
\Gamma_{\mathrm{FAS}} &\ge \frac{|g_{\mathrm{FAS}}|^2}{\left(U-1\right)\left(\sigma^2+\left(U-2\right)\frac{\pi}{4}\sigma^2\bar{\rho}\right)+\sigma^2_{\eta}}.
\end{aligned}
\end{equation}
In the following, the SINR lower-bound for FBL-FAS will be adopted for OP calculation. Also, from \eqref{eq.OP_def}, we get
\begin{equation}
\begin{aligned}
&p^{\mathrm{FBL\text{-}FAS}}_{\mathrm{out}}(\gamma^{\mathrm{FAS}}_{\mathrm{th}})\\
&=P\left\{\Gamma_{\mathrm{FAS}}\le 2^{R_c}-1\right\}=P\left\{|g_{\mathrm{FAS}}|\le \gamma^{\mathrm{FAS}}_{\mathrm{th}}\right\},
\end{aligned}
\end{equation}
 where $\gamma^{\mathrm{FAS}}_{\mathrm{th}}=\sqrt{ \left(2^{R_c}-1\right)\left(\left(U-1\right)\left(\sigma^2+\left(U-2\right)\frac{\pi}{4}\sigma^2\bar{\rho}\right)+\sigma^2_{\eta}\right)}$ is the outage threshold. Eventually, utilizing the joint CDF in \eqref{joint CDF} completes the proof.

\section{Proof of MRC-OP for $L$-FPA Systems}\label{appen-i}
Consider a conventional system with $L$ independent FPAs employing MRC. The received signal matrix $\boldsymbol{Y} \in \mathbb{C}^{M \times L}$ over a block of $M$ channel uses is given by
\begin{equation}
\boldsymbol{Y} = \boldsymbol{x}_i \boldsymbol{g}_i + \sum_{u \neq i}^{U} \boldsymbol{x}_u \boldsymbol{g}_u + \boldsymbol{Z},
\end{equation}
where $\boldsymbol{x}_i, \boldsymbol{x}_u \in \mathbb{C}^{M \times 1}$ are the transmitted codewords normalized such that $\|\boldsymbol{x}\|_2^2=1$, channel vectors $\boldsymbol{g}_i, \boldsymbol{g}_u \in \mathbb{C}^{1 \times L}$ are the quasi-static channel vectors with independent entries following $\mathcal{CN}(0, \sigma^2)$ and $\boldsymbol{Z} \in \mathbb{C}^{M \times L}$ is the additive noise matrix with entries following $\mathcal{CN}(0, \frac{\sigma^2_{\eta}}{M})$. The MRC receiver applies the linear combining vector $\boldsymbol{w} = \boldsymbol{g}_i^{\mathrm{H}}$ to the received signal, yielding the combined signal vector $\widehat{\boldsymbol{y}} = \boldsymbol{Y}\boldsymbol{g}_i^{\mathrm{H}} \in \mathbb{C}^{M \times 1}$:
\begin{equation}
	\widehat{\boldsymbol{y}} = \boldsymbol{x}_i \|\boldsymbol{g}_i\|^2 + \sum_{u \neq i}^{U} \boldsymbol{x}_u (\boldsymbol{g}_u \boldsymbol{g}_i^{\mathrm{H}}) + \boldsymbol{Z}\boldsymbol{g}_i^{\mathrm{H}}.
\end{equation}
Similar to the SINR definition in \eqref{eq:SINR_def_robust}, the instantaneous SINR, $\Gamma_{\mathrm{MRC}}$, is derived by analyzing the power of the signal, interference, and noise components conditioned on $\boldsymbol{g}_i$:
\begin{itemize}
\item $P_S$: The average power of the desired signal component is
\begin{equation}\label{eq.P_S_L}
P_S = \left| \|\boldsymbol{g}_i\|^2 \right|^2 = \|\boldsymbol{g}_i\|^4.
\end{equation}
\item $P_I$: The interference term is a sum of random variables. Similar to the derivations in \eqref{eq:Pi_approx}, we could approximate the average power of the interference component as
\begin{equation}\label{eq._P_I_L}
\begin{aligned}
P_I &= \left| \sum_{u \neq i}^{U} \boldsymbol{g}_u \boldsymbol{g}_i^{\mathrm{H}} \right|^2 \\
&\le \sum_{u \neq i}^{U} |\boldsymbol{g}_u \boldsymbol{g}_i^{\mathrm{H}}|^2 + \sum_{u \neq i}^{U} \sum_{v \neq i\atop v \neq u}^{U} |\boldsymbol{g}_u \boldsymbol{g}_i^{\mathrm{H}}| |\boldsymbol{g}_v \boldsymbol{g}_i^{\mathrm{H}}| \rho_{u,v},
\end{aligned}
\end{equation}
where we approximate the codebook correlation $| \boldsymbol{x}_u^{\mathrm{H}} \boldsymbol{x}_v |$ by the average correlation $\bar{\rho}$. We then approximate the channel terms by their statistical expectations conditioned on $\boldsymbol{g}_i$. Specifically, for the complex Gaussian projection $\boldsymbol{g}_u \boldsymbol{g}_i^{\mathrm{H}} \sim \mathcal{CN}(0, \|\boldsymbol{g}_i\|^2 \sigma^2)$, we have
\begin{subequations}\label{eq:channel_terms}
\begin{align}
\mathrm{E}_{\boldsymbol{g}_i}\left[ |\boldsymbol{g}_u \boldsymbol{g}_i^{\mathrm{H}}|^2  \right]& = \|\boldsymbol{g}_i\|^2 \sigma^2,\\
\mathrm{E}_{\boldsymbol{g}_i}\left[ |\boldsymbol{g}_u \boldsymbol{g}_i^{\mathrm{H}}| \right] &= \sqrt{\|\boldsymbol{g}_i\|^2 \sigma^2} \frac{\sqrt{\pi}}{2} = \|\boldsymbol{g}_i\| \sigma \frac{\sqrt{\pi}}{2}.
\end{align}
\end{subequations}
Substituting \eqref{eq:channel_terms} into \eqref{eq._P_I_L} yields the approximated interference power
\begin{equation}\label{eq.P_I_L_2}
\begin{aligned}
P_I &\le (U-1)\|\boldsymbol{g}_i\|^2 \sigma^2 + (U-1)(U-2) \left( \|\boldsymbol{g}_i\| \sigma \frac{\sqrt{\pi}}{2} \right)^2 \bar{\rho} \\
&= \|\boldsymbol{g}_i\|^2 \sigma^2 \left( (U-1) + (U-1)(U-2) \frac{\pi}{4} \bar{\rho} \right).
\end{aligned}
\end{equation}
\item $P_N$: The combined noise vector is $\tilde{\boldsymbol{z}} = \boldsymbol{Z}\boldsymbol{g}_i^{\mathrm{H}}$. Given that the entries of $\boldsymbol{Z}$ have variance $\sigma^2_{\eta}/M$. Similar to \eqref{eq:Pn_robust}, the average power of the combined noise is found as
\begin{equation}\label{eq.P_N_L}
P_N = \|\boldsymbol{g}_i\|^2 \sigma^2_{\eta}.
\end{equation}
\end{itemize}
Substituting \eqref{eq.P_S_L}, \eqref{eq.P_I_L_2} and \eqref{eq.P_N_L} into the SINR definition $\Gamma = P_S / (P_I + P_N)$, the term $\|\boldsymbol{g}_i\|^2$ cancels out from the numerator and denominator, resulting in
\begin{equation}\label{eq:MRC_SINR_Final}
\Gamma_{\mathrm{MRC}} \ge \frac{\|\boldsymbol{g}_i\|^2}{\sigma^2 \left( U-1 + (U-1)(U-2)\frac{\pi}{4}\bar{\rho} \right) + \sigma^2_{\eta}}.
\end{equation}
Subsequently, the OP is derived as
\begin{equation}
\begin{aligned}
p^{\mathrm{MRC}}_{\mathrm{out}}(\gamma^{\mathrm{MRC}}_{\mathrm{th}}) &= P\left\{ \log_2(1 + \Gamma_{\mathrm{MRC}}) \le R_c \right\}\\
& = P\left\{ \|\boldsymbol{g}_i\|^2 \le \gamma^{\mathrm{MRC}}_{\mathrm{th}} \right\},
\end{aligned}
\end{equation}
where the outage threshold for MRC $L$-FPA system is
\begin{equation}
\gamma^{\mathrm{MRC}}_{\mathrm{th}}=\left (\left(U-1 \right)\left( \sigma^2  + \sigma^2 (U-2)\frac{\pi}{4}\bar{\rho} \right) + \sigma^2_{\eta}\right)\left(2^{R_c}-1\right).
\end{equation}

The random variable $X = \|\boldsymbol{g}_i\|^2 = \sum_{l=1}^{L} |g_{i,l}|^2$ follows an Erlang distribution, a special case of the Gamma distribution, with shape parameter $L$ and scale parameter $\sigma^2$, whose CDF is given by $F_X(x) = 1 - e^{-x/\sigma^2} \sum_{k=0}^{L-1} \frac{1}{k!} (x/\sigma^2)^k$. Therefore, the OP is obtained as
\begin{equation}\label{eq:OP_Closed_Form}
p^{\mathrm{MRC}}_{\mathrm{out}} = 1 - \exp\left( - \frac{\gamma^{\mathrm{MRC}}_{\mathrm{th}}}{\sigma^2} \right) \sum_{k=0}^{L-1} \frac{1}{k!} \left( \frac{\gamma^{\mathrm{MRC}}_{\mathrm{th}}}{\sigma^2} \right)^k,
\end{equation}
which completes the proof.

\section{Proof of Identities in Theorem~\ref{appen-e}}\label{ID_Proof}
   \begin{enumerate}
   \item{\em Identity 1:} 
   \begin{equation*}
   	\frac{\partial}{\partial d}Q_1(c,d)=-d e^{\!\left(-\frac{c^2+d^2}{2}\right)}I_0(cd).
   \end{equation*}
   \begin{IEEEproof}
Let $f(x;c)=xe^{\!\left(-\frac{x^2+c^2}{2}\right)}I_0(cx)$. Then, $Q_1(c,d)=\int_d^{\infty} f(x;c)\,\mathrm{d}x$. By the Leibniz integral rule for an integral with variable lower limit, $\frac{\partial}{\partial d}\int_d^\infty f(x;c)\,\mathrm{d}x=-f(d;c).$ Hence,
$\frac{\partial}{\partial d}Q_1(c,d)=-\left[de^{\!\left(-\frac{d^2+c^2}{2}\right)}I_0(cd)\right]$,
which completes the proof.
   	\end{IEEEproof}
\item {\em Identity 2:}
 \begin{equation*}
 	\frac{\partial}{\partial r} Q_1(a \sqrt{t}, b r) = -b^2 e^{\left( -\frac{a^2 t + b^2 r^2}{2} \right)} I_0(a \sqrt{t} b r).
 \end{equation*}
\begin{IEEEproof}
Define $A=a\sqrt{t},~D=br.$ Then, $Q_1(a\sqrt{t},br)=Q_1(A,D)$. Applying the chain rule yields $ \frac{\partial}{\partial r}Q_1(a\sqrt{t},br)=\frac{\partial Q_1(A,D)}{\partial D}\frac{\partial D}{\partial r}$. From Identity 1, we have $\frac{\partial Q_1(A,D)}{\partial D}=-D\exp\!\left(-\frac{A^2+D^2}{2}\right)I_0(AD),$ and clearly $
\frac{\partial D}{\partial r}=b.$ Therefore,
\begin{equation}
	\frac{\partial}{\partial r}Q_1(a\sqrt{t},br) = \left[-De^{\!\left(-\frac{A^2+D^2}{2}\right)}I_0(AD)\right]b.
\end{equation}
Substituting $D=br$ and $A=a\sqrt{t}$ gives $
\frac{\partial}{\partial r}Q_1(a\sqrt{t},br)
=
-b^2r\,e^{\!\left(-\frac{a^2t+b^2r^2}{2}\right)}I_0(a\sqrt{t}br),$
which completes the proof.
\end{IEEEproof}
   \end{enumerate}
\end{appendices}

\balance

\end{document}